\documentclass[superscriptaddress,twocolumn,secnumarabic,
amssymb,amsmath,nobibnotes,aps,prd,showpacs,nofootinbib]{revtex4}%
\usepackage{graphicx}
\usepackage{epsf}
\usepackage{bm}
\usepackage{amsmath}
\usepackage{amsfonts}
\usepackage{amssymb}
\usepackage{epstopdf}
\usepackage{natbib}
\usepackage{color}%
\providecommand{\U}[1]{\protect\rule{.1in}{.1in}}
\newcommand{\be}{\begin{equation}}
\newcommand{\ee}{\end{equation}}

\newcommand{\mincir}{\raise
-3.truept\hbox{\rlap{\hbox{$\sim$}}\raise4.truept\hbox{$<$}\ }}
\newcommand{\magcir}{\raise
-3.truept\hbox{\rlap{\hbox{$\sim$}}\raise4.truept\hbox{$>$}\ }}

\usepackage{color}

\begin{document}

\title{Interacting dark energy with time varying equation of state and the $H_0$ tension}

\author{Weiqiang Yang}
\email{d11102004@163.com}
\affiliation{Department of Physics, Liaoning Normal University, Dalian, 116029, P. R. China.}

\author{Ankan Mukherjee}
\email{ankanju@iisermohali.ac.in}
\affiliation{Department of Physical Sciences, Indian Institute of Science Education and Research Mohali, Sector 81, Mohali, Punjab 140306, India.}

\author{Eleonora Di Valentino}
\email{eleonora.divalentino@manchester.ac.uk}
\affiliation{Jodrell Bank Center for Astrophysics, School of Physics and Astronomy, University of Manchester, Oxford Road, Manchester, M13 9PL, UK.}

\author{Supriya Pan}
\email{supriya.maths@presiuniv.ac.in}
\affiliation{Department of Mathematics, Presidency University, 86/1 College Street, Kolkata 700073, India.}

\pacs{98.80.-k, 95.36.+x, 95.35.+d, 98.80.Es}

\begin{abstract}
Almost in all interacting dark energy models present in the literature, the stability of the model becomes potentially sensitive to the dark energy equation of state parameter $w_x$, and a singularity arises at `$w_x = -1$'. Thus, it becomes mandatory to test the stability of the model into two separate regions, namely, for quintessence and phantom. This essentially brings in a discontinuity into the parameters space for $w_x$. Such discontinuity can be removed with some specific choices of the interaction or coupling function. In the present work we choose one particular coupling between dark matter and dark energy which can successfully remove such instability and we allow a dynamical dark energy equation of state parameter instead of the constant one. In particular, considering a dynamical dark energy equation of state with only one free parameter $w_0$, representing the current value of the dark energy equation of state, we confront the interacting scenario with several observational datasets. The results show that the present cosmological data allow an interaction in the dark sector, in agreement with some latest claims by several authors, and additionally, a phantom behaviour in the dark energy equation of state is suggested at present. Moreover, for this case the tension on $H_0$ is clearly released. As a final remark, we mention that according to the Bayesian analysis, $\Lambda$-cold dark matter ($\Lambda$CDM) is always favored over this interacting dark energy model. 

\end{abstract}

\maketitle

\section{Introduction}

The accelerated expansion of the universe still remains as an enigma for cosmologists. It was first discovered in late nineties from the observations of nearby Supernovae Type Ia \cite{snia1,snia2}. Further observations, like the Baryon Oscillation Spectroscopic
Survey (BOSS) \cite{Schlegel:2009hj}, the continuation of Supernova cosmology project \cite{Suzuki:2011hu}, the Dark Energy
Survey \cite{Crocce:2015xpb}, the mapping of the universe from the multi wavelength observations of the Sloan
Digital Sky Survey (SDSS) \cite{Alam:2015mbd}, the observation of cosmic microwave background (CMB) from
WMAP~\cite{Hinshaw:2012aka}, Planck \cite{Ade:2015xua,Aghanim:2018eyx}  and several other observations have strongly confirmed the accelerated expansion of the universe at recent time. There are different theoretical prescriptions in the literature to explain this late-time cosmic acceleration. The most popularly accepted one is the assumption of the existence of an exotic component in the energy budget of the universe. The exotic component, dubbed as dark energy, is responsible for the alleged accelerated expansion due to its negative pressure. Observations suggest that dark energy contributes around $70\%$ to the total energy density of the universe \cite{Aghanim:2018eyx}. The rest of the contribution is predominated by another exotic component, called the dark matter (roughly around $26\%$) \cite{Aghanim:2018eyx}. The fundamental difference between dark matter and ordinary {\bf non-exotic} baryonic matter is  that the dark matter does not have any electromagnetic, strong or week interactions like baryonic matter, though they have similar gravitational interaction.

The present work is mainly focused on the interaction between dark energy and the dark matter. There are existing models in the literature where independent conservation of dark energy and dark matter has been assumed, see the details here \cite{Copeland:2006wr}. On the other hand, models which allow the interaction between these two dark components, are also well consistent with cosmological observations \cite{Salvatelli:2014zta, Yang:2017yme, Nunes:2016dlj}. In fact, interacting and non-interacting dark energy models are not sharply distinguishable from the present available observational data. Though the non-interacting dark energy models are well enough to explain the observed cosmological scenario, they suffers from certain theoretical issues. The allowance of an interaction between dark energy and dark matter was originally motivated to provide with an explanation to the extremely small value of the cosmological constant \cite{Wetterich-ide1}  and later on it was found that an interaction between dark matter and dark energy is able to solve the well known cosmic coincidence problem \cite{Amendola-ide1,Amendola-ide2,Pavon:2005yx,delCampo:2008sr,delCampo:2008jx}. The developments in  observational cosmology successively fueled the investigations in this topic with many interesting outcomes, see  \cite{Salvatelli:2014zta,Yang:2014gza,Yang:2014hea,Nunes:2016dlj,Kumar:2016zpg,vandeBruck:2016hpz,
Yang:2016evp, Caprini:2016qxs, Cai:2017yww, Kumar:2017bpv,Pan:2017ent,Yang:2018ubt,Yang:2018pej,Yang:2018euj,Yang:2018xlt} for different aspects of interacting dark energy from recent observations. Additionally, several theoretical developments have also enriched the literature of interacting dark matter and dark energy scenarios, see for instance \cite{Bolotin:2013jpa,Pan:2014afa, Shahalam:2015sja, Shahalam:2017fqt}.

Interacting dark energy models, also known as coupled dark energy models, are mainly designed through the parametrizations of the interaction function that appears in the conservation equations of dark matter and dark energy. 
In the present work, we have considered a specific interaction scenario, where one of the interacting fluids, namely, dark energy
has a time dependent equation of state ($w_x$). We refer to some recent works where the interaction scenarios with time varying dark energy equation of state have been investigated \cite{Yang:2017yme,Pan:2016ngu,Sharov:2017iue,Wang:2014iua,Pan:2012ki,He:2008tn,Yang:2017ccc}. The dark energy equation of state has an essential role in this context because the stability of the interaction model at large scales of the universe is highly sensitive to it. Whenever the dark energy state parameter crosses the phantom divide line `$w_x = -1$', the modified perturbation equations in presence of an interaction/coupling between the dark components, become undefined leading to a singularity at $w_x=-1$.
Thus, in order to confront the interactions/couplings with the observational data, two separate regions, namely,  
$w_x> -1$ and $w_x < -1$ are considered. 
However, such problems can be dodged with some new kinds of  
interaction models, recently explored 
in \cite{Yang:2017zjs,Yang:2017ccc}. Additionally, some recent astronomical observations report some  interesting and overwhelming issues on the coupling between  
the dark components. Precisely, according to the observational data, a nonzero coupling in the dark sectors is allowed \cite{Salvatelli:2014zta, Nunes:2016dlj,Kumar:2016zpg, vandeBruck:2016hpz, Yang:2017yme, Kumar:2017dnp}, although the coupling is small, but a small deviation from the standard $\Lambda$-cosmology is not completely ruled out. 
The tension on the Hubble constant $H_0$
appearing from the local and global measurements are found to 
be assuaged in presence of an interaction between dark matter and dark energy  \cite{DiValentino:2017iww, Kumar:2016zpg,Yang:2017zjs, Bhattacharyya:2018fwb}.  The inclusion of the coupling between the dark components may also push the dark energy equation of state to go beyond the cosmological constant limit `$w_x = -1$' \cite{Yang:2017zjs,Yang:2017ccc,Nunes:2016dlj, Yang:2017yme, Pan:2016ngu, Sharov:2017iue}. 
Therefore, it is quite certain that the coupling in the dark sectors still remains 
as an attracting field for further investigations. 
Now, compared to the interacting scenarios with constant dark energy equation of 
state \cite{DiValentino:2017iww, Kumar:2016zpg,Nunes:2016dlj, Kumar:2016zpg,Yang:2017zjs}, the same with dynamical equation of state has not been much  explored except of some minimal investigations \cite{Yang:2017yme, Pan:2016ngu, Sharov:2017iue, Wang:2014iua, Pan:2012ki,Yang:2017ccc}. Thus, in this work we perform a systematic analysis for dynamical dark energy coupled to dark matter.  In fact, the equation of state for dark energy evolving with time is most preferred scenario as found in several analyses 
\cite{DiValentino:2017zyq,Zhao:2017cud,DiValentino:2017gzb, Sola:2017lxc,Gomez-Valent:2018nib}.
In the present analysis, we focus on a specific parameterization of the dark energy equation of state, namely, 
a one parameter dark energy model \cite{Gong:2005de}  and we 
constrain the coupling strength of the interaction function along with other free parameters of the interacting model as well.

The present work has  been organized in the following way. In section \ref{sec-2}, we describe the basic equations of the interaction models at the background and perturbative levels as well as we introduce the specific interaction model that has been studied in the present context. In section \ref{sec-data}, we first describe the observational data to constrain the interaction scenarios and then we describe the results of the analysis in subsections \ref{sec:ide1} and \ref{sec-cmb+matter}. A Bayesian analysis for statistical model selection through the calculation of Bayesian evidence has been discussed in subsection \ref{sec-bayesian}. Finally, we close the work with a brief summary in section \ref{sec-conclu}.

\section{Interacting dark fluids at the background and perturbative levels}
\label{sec-2}

In the cosmological length scale, the geometry of the universe is best described by the 
Friedman-Lema\^{i}tre-Robertson-Walker (FLRW) line element. Thus, in this work  we assume the same line element which takes the form 
\begin{eqnarray}
ds^2 = -dt^2 + a^2 (t) \left[\frac{dr^2}{1-Kr^2} + \left( d\theta ^2 + \sin^2 \theta d\phi^2\right)  \right],
\end{eqnarray}
where $a(t)$ is the scale factor of the universe and $K$ is its curvature scalar. For $K=0, -1, +1$, a spatially flat, open or a closed universe is respectively described. In the present work 
we consider the spatial flatness of the universe, that means we set $K =0$, throughout the present work. Further, we assume that the gravity sector of the universe follows the Einstein's general relativity where additionally, (i) the matter sector is minimally coupled to gravity and (ii) the total energy density of the universe is shared by four components, namely, radiation, baryons, pressureless dark matter and  a dark energy fluid where only dark matter and dark energy fluids are coupled to each other while the rest two fluids are conserved separately.  The conservation equations for dark matter with zero pressure, i.e., cold dark matter (CDM) and dark energy (DE) with dynamical equation of state, $w_x \equiv p_x/\rho_x$, can be given as 

\begin{eqnarray}\label{cons-dm}
\dot{\rho}_c + 3 H \rho_c = -Q,
\end{eqnarray} 
and 
\begin{eqnarray}\label{cons-de}
\dot{\rho}_x + 3 H (1+w_x) \rho_x = Q,
\end{eqnarray} 
where $H \equiv \dot{a}/a$ is the Hubble rate of this FLRW universe and in equations (\ref{cons-dm}), (\ref{cons-de}), the new quantity $Q$ describes  the flow of energy between the dark sectors (i.e., CDM and DE), known as the interaction function. The
algebraic constraint on the dynamics of the universe is the Friedmann equation,
\begin{eqnarray}\label{friedmann}
H^2 = \frac{8 \pi G}{3} \left( \rho_r + \rho_b + \rho_c + \rho_x  \right),
\end{eqnarray}
where $G$ is the Newton's gravitational constant. The equation (\ref{friedmann}) together with the conservation equations (\ref{cons-dm}) and (\ref{cons-de})
can in principle determine the entire dynamics of the universe, once the interaction function $Q$ 
is prescribed. Usually, several choices for $Q$ can be made, however, in this work we are interested on the choice of the interaction functions that are able to produce stable perturbations on the large scales of our Universe. Since the structure formation is a very important issue to understand the dynamics of the universe, thus, it is mandatory to focus on the perturbations equations that are modified in presence of any arbitrary coupling between dark matter and dark energy.  

\begin{figure*}
\includegraphics[width=0.48\textwidth]{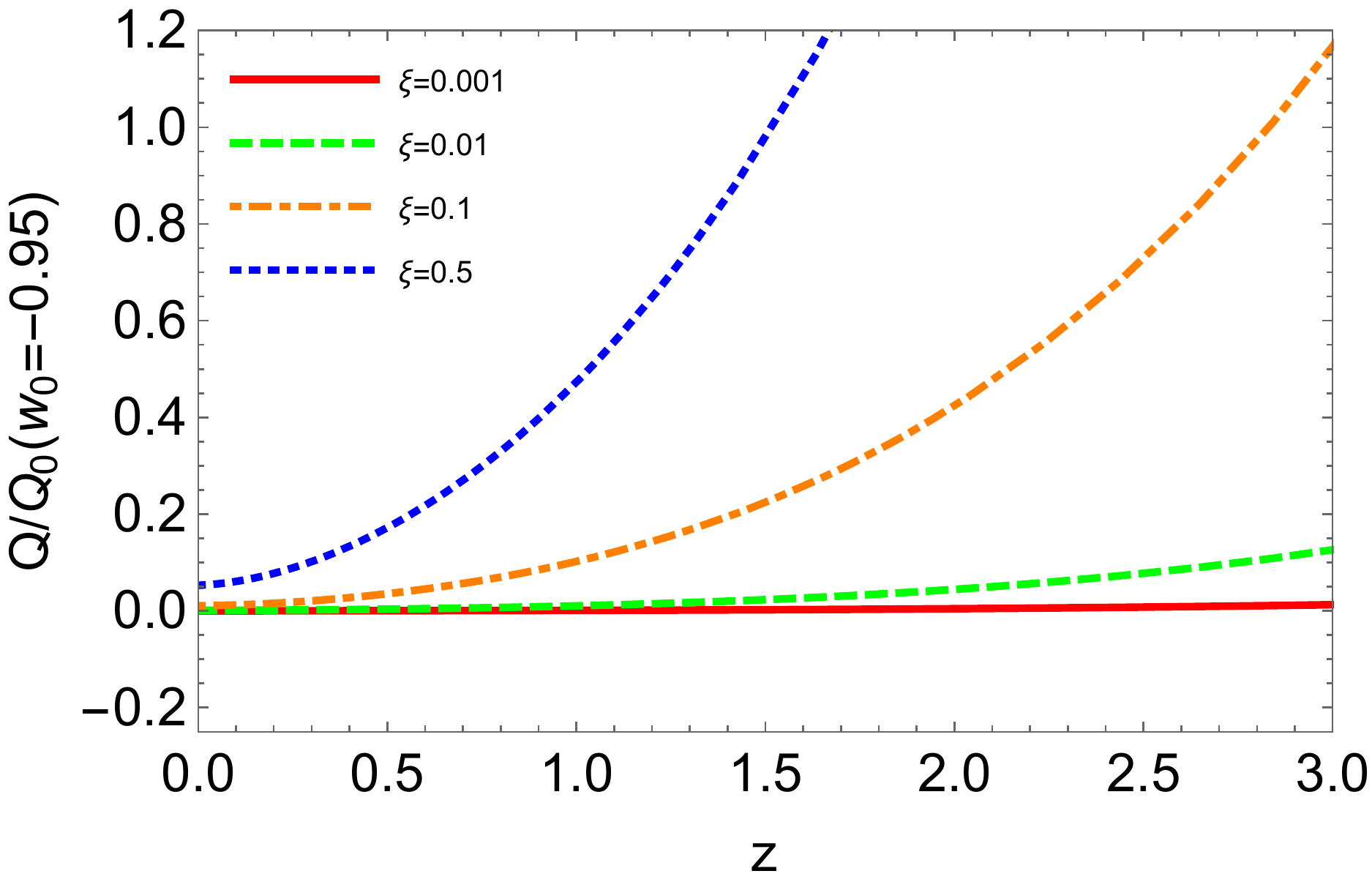}
\includegraphics[width=0.48\textwidth]{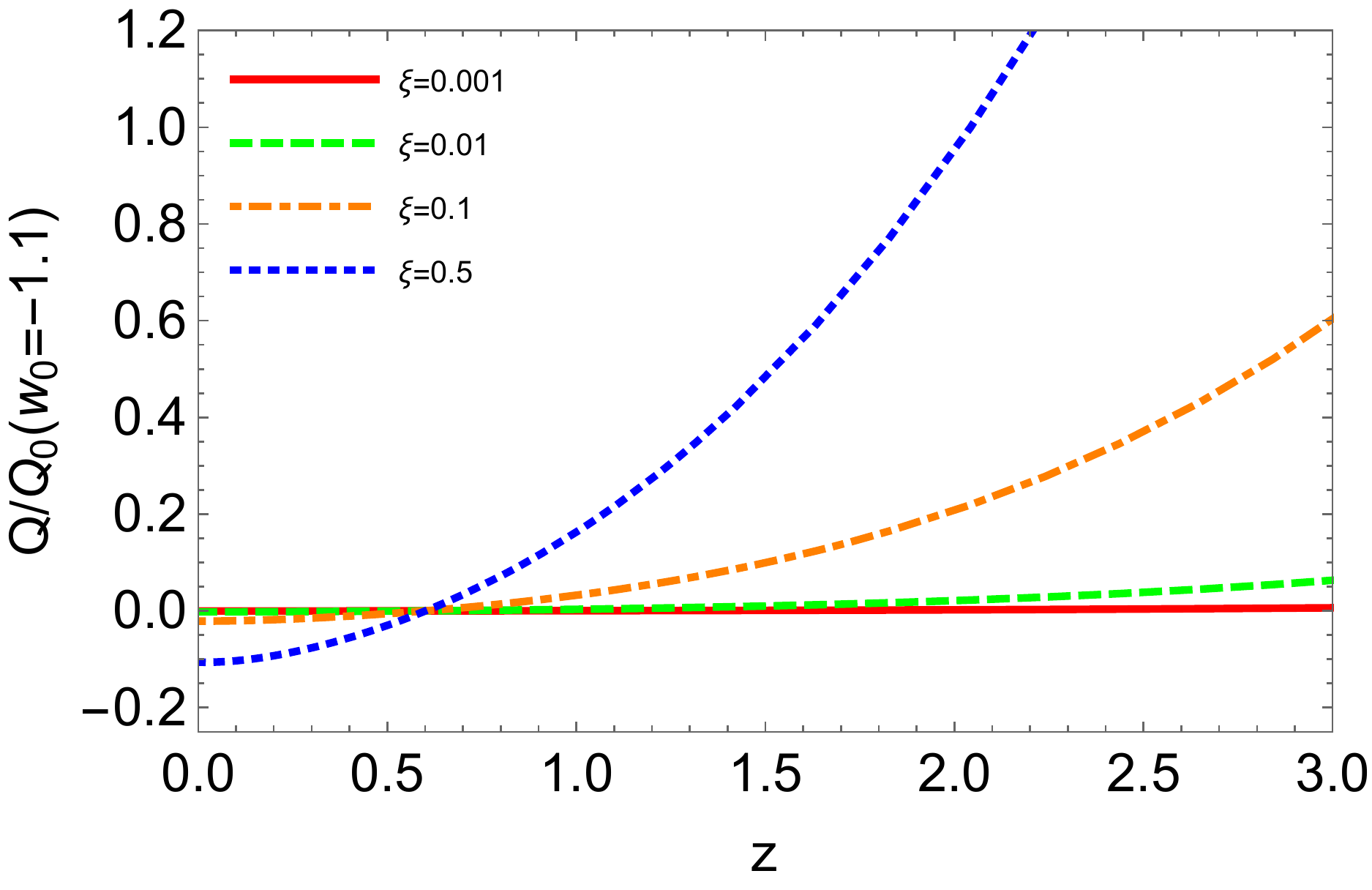}
\caption{Qualitative evolution of the interaction model $Q = 3 H \xi (1+w_x) \rho_{x}$ where $w_x$ is given in eqn. (\ref{eq:ZG-II}) has been shown for some specific choices of the coupling parameter, $\xi$. In the left panel we exhibit the behaviour of the interaction function for quintessence kind of dark energy, i.e., $w_0 > -1$ (in particular, we set $w_0 = -0.95$) while the right panel depicts the evolution of the interaction function but for phantom dark energy state parameter, that means for $w_0 < -1$ (in particular, $w_0 = -1.1$). Let us note that $Q_0 = H_0 \rho_{tot,0} = 3 H_0^3 /(8 \pi G)$ where $\rho_{tot} = \left( \rho _{r}+\rho_{b}+\rho _{c}+\rho _{x}\right)$, is the total energy density of the universe and $\rho_{tot,0} = \rho_{tot} (z=0)$.  }
\label{fig:Q-xi}
\end{figure*}

\begin{figure*}
\includegraphics[width=0.5\textwidth]{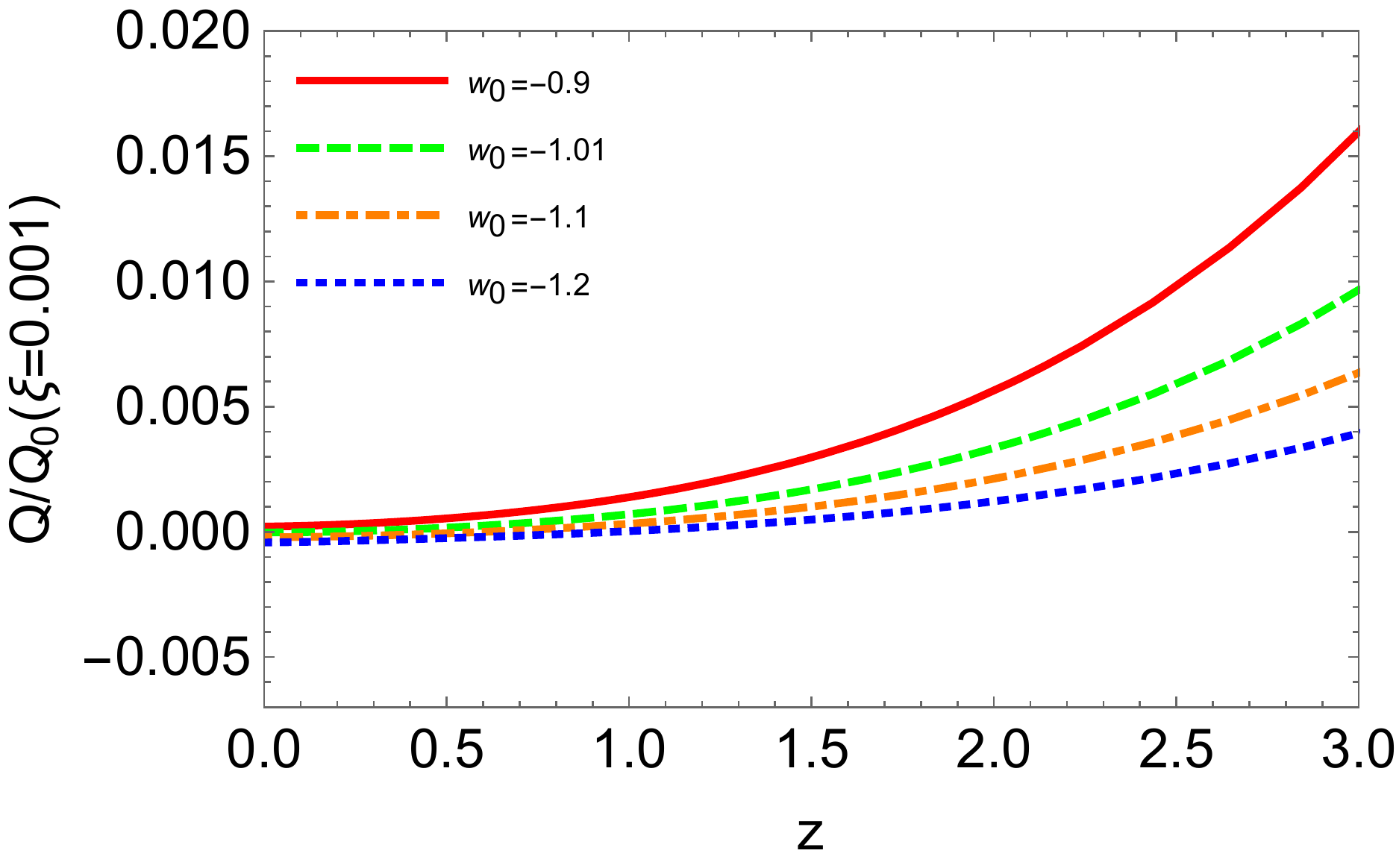}
\includegraphics[width=0.47\textwidth]{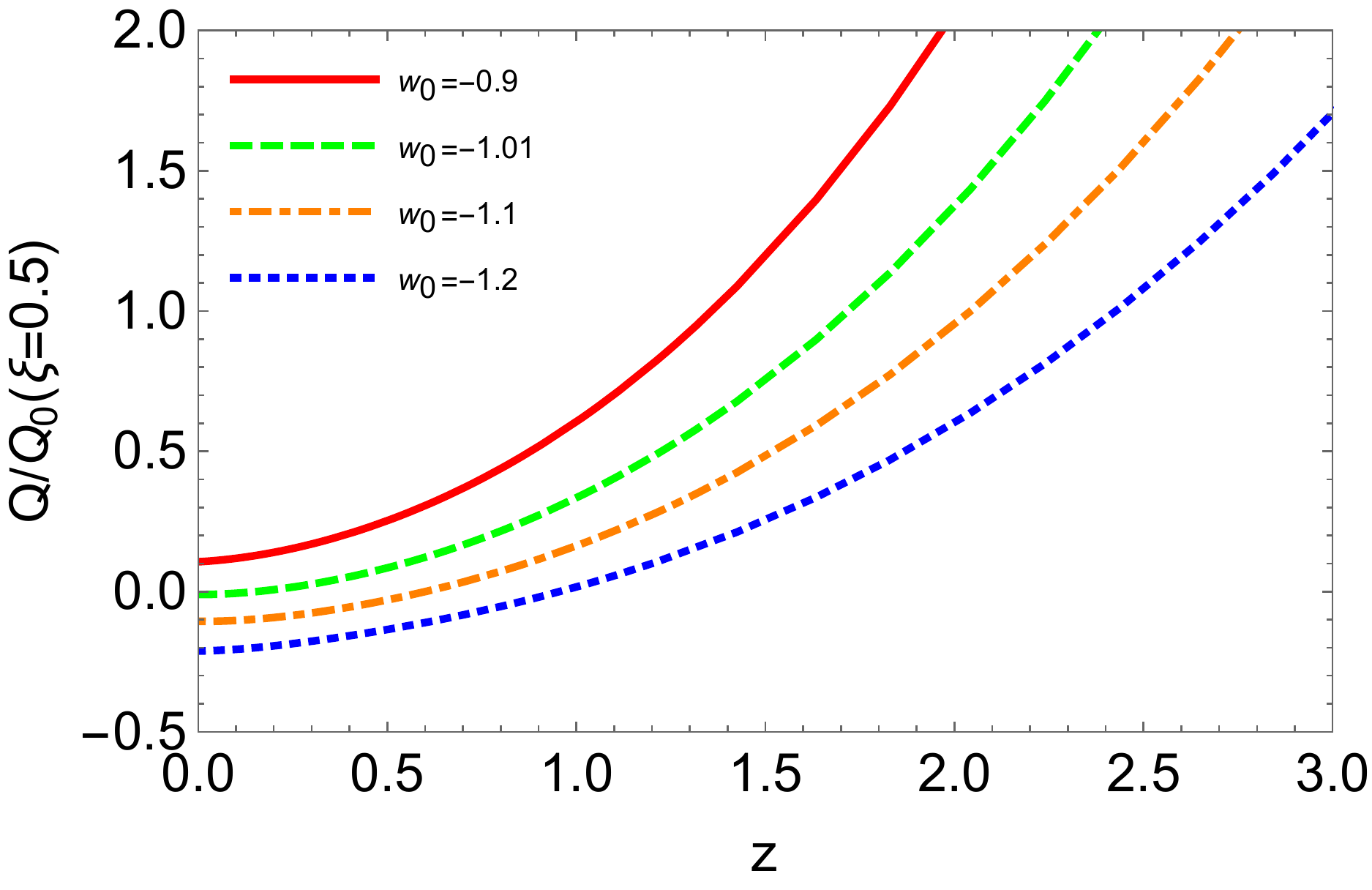}
\caption{The figure depicts the evolution of the interaction function for different values of $w_0$ with some fixed coupling strengths. The left panel portrays the evolution of $Q$ for different values of $w_0$ with a fixed and low coupling strength $\xi = 0.001$ while on the other hand, the right panel shows the same evolution but for a large coupling strength $\xi = 0.5$. We note that $Q_0$ has similar meaning as described for Fig.  \ref{fig:Q-xi}. }
\label{fig:Q-w}
\end{figure*}

In what follows, we consider the perturbed FLRW metric given by  \cite{Mukhanov, Ma:1995ey, Malik:2008im}
\begin{eqnarray}
ds^{2}=a^{2}(\tau )\Bigg[-(1+2\phi )d\tau ^{2}+2\partial _{i}Bd\tau dx^{i}\notag\\+
\Bigl((1-2\psi )\delta _{ij}+2\partial _{i}\partial _{j}E\Bigr)dx^{i}dx^{j}%
\Bigg],
\end{eqnarray}%
where by $\tau$ we mean the conformal time; 
$\phi $, $B$, $\psi $ and $E$ are the 
the gauge-dependent scalar perturbation quantities. 
For the above metric, one can calculate the field 
equations as  \cite{Majerotto:2009np, Valiviita:2008iv, Clemson:2011an}
\begin{equation*}
\nabla _{\nu }T_{A}^{\mu \nu }=Q_{A}^{\mu },~~~~\sum\limits_{\mathrm{A}}{%
Q_{A}^{\mu }}=0,
\end{equation*}%
where $A$ has been used to mean any fluid either dark matter or dark energy. For $A = c$, we mean CDM while $A =x$ means the DE fluid. 
The quantity
$Q_{A}^{\mu }$ takes the form  
\begin{eqnarray}
Q_{A}^{\mu }=(Q_{A}+\delta Q_{A})u^{\mu }+a^{-1}(0,\partial
^{i}f_{A}), 
\end{eqnarray}
relative to the four-velocity $u^{\mu }$ where $Q_A$ is the background energy
transfer (i.e., $Q_A = Q$) and $f_A$ is the momentum transfer potential. 
For simplicity, we assume that momentum transfer potential is 
zero in the rest frame of the dark matter 
\cite{Majerotto:2009np, Valiviita:2008iv, Clemson:2011an} which directs 
$k^{2}f_{A}=Q_{A}(\theta -\theta _{c})$ where $k$ is the wave number and 
$\theta = \theta_{\mu}^{\mu}$, $\theta_c$ are respectively 
the volume expansion scalar of the total fluid and the volume expansion scalar for the CDM fluid. Now, introducing the
density perturbations for the fluid `$A$' as   
$\delta _{A}=\delta \rho _{A}/\rho
_{A}$ and considering no anisotropic stress in the system, 
the density and velocity perturbations for the dark fluids in the 
synchronous gauge, that means $\phi =B=0$, $\psi =\eta $, and $k^{2}E=-h/2-3\eta $, where $h$ and $\eta$ are the metric perturbations (see \cite{Ma:1995ey} for details),
can be written as 
\begin{widetext}
\begin{eqnarray}
\delta _{x}^{\prime } &=&-(1+w_{x})\left( \theta _{x}+\frac{h^{\prime }}{2}%
\right) -3\mathcal{H}(c_{sx}^{2}-w_{x})\left[ \delta _{x}+3\mathcal{H}%
(1+w_{x})\frac{\theta _{x}}{k^{2}}\right] -3\mathcal{H}w_{x}^{\prime }\frac{%
\theta _{x}}{k^{2}}  \notag \\
&+&\frac{aQ}{\rho _{x}}\left[ -\delta _{x}+\frac{\delta Q}{Q}+3\mathcal{H}%
(c_{sx}^{2}-w_{x})\frac{\theta _{x}}{k^{2}}\right] , \\
\theta _{x}^{\prime } &=&-\mathcal{H}(1-3c_{sx}^{2})\theta _{x}+\frac{%
c_{sx}^{2}}{(1+w_{x})}k^{2}\delta _{x}+\frac{aQ}{\rho _{x}}\left[ \frac{%
\theta _{c}-(1+c_{sx}^{2})\theta _{x}}{1+w_{x}}\right] , \\
\delta _{c}^{\prime } &=&-\left( \theta _{c}+\frac{h^{\prime }}{2}\right) +%
\frac{aQ}{\rho _{c}}\left( \delta _{c}-\frac{\delta Q}{Q}\right) , \\
\theta _{c}^{\prime } &=&-\mathcal{H}\theta _{c},  \label{eq:perturbation}
\end{eqnarray}
\end{widetext}
where $\mathcal{H}$ is the conformal Hubble rate and the quantity $\delta Q/Q$ actually includes the perturbations for the 
Hubble rate $\delta H$ (recall that $\mathcal{H}=aH$). Using $\delta H$, 
the  gauge invariant equations for the coupled dark fluids 
can be easily found \cite{Gavela:2009cy}. Thus, in this analysis, 
we consider the perturbation of the Hubble expansion rate since the
total expansion rate includes both background and perturbation.
Now, the stability of the model depends on the pressure perturbations
for dark energy which is also dependent on the interaction function 
through the relation, 
\begin{eqnarray}
\delta p_{x} &=& c_{sx}^{2}\delta \rho _{x}-(c_{sx}^{2}-c_{ax}^{2})\rho
_{x}^{\prime }\frac{\theta _{x}}{k^{2}}\notag\\
&=&c_{sx}^{2}\delta \rho _{x}+3\mathcal{H}\rho
_{x}(1+w_{x})(c_{sx}^{2}-c_{ax}^{2})(1+d)\frac{\theta _{x}}{k^{2}},\notag
\label{eq:deltap}
\end{eqnarray}
where the parameter $d$, named as doom factor, is expressed as
\begin{equation}\label{doom}
d\equiv -aQ/[3\mathcal{H}\rho _{x}(1+w_{x})],
\end{equation}%
It ensures the stability 
of any interaction model  for $d\leq 0$ \cite{Gavela:2009cy}. Thus, using the 
expression for the doom factor (\ref{doom}), for 
any interaction model, one can 
find the conditions for stability of the interaction model.
In particular, for the 
usual models $Q = \xi H \bar{Q}$ (where $\bar{Q} >0$)
one can find that, the 
model could lead to stable perturbations at large scale if 
$\xi \geq 0$~and~$(1+w_{x})>0$ or $\xi \leq 0$~and$~(1+w_{x})<0$. 
So, clearly there is a jump of the equation of state $w_x$ at
`$-1$'. We mention that if we simply consider the interacting 
cosmological constant scenario, then the governing equations
become simple from the very beginning and the treatment is no 
longer the same for any arbitrary $w_x \neq -1$. However, for 
any arbitrary $w_x$, if we need to constrain the dark 
energy equation of state, then we cannot take the prior 
of $w_x \in [a, b]$ ($a, b \in \mathbb{R}$) where $-1$ is included in this closed
interval. We must have to take the intervals $(-1, b]$ or $[a, -1)$. Thus, certainly,
it is clear that some information is basically lost during the analysis. Such problem 
can be removed if we simply transform the interaction as $Q \rightarrow (1+w_x) Q$, that is, if we include a term $(1+w_x)$ from outside into the interaction function.  However, such phenomenological construction can be viewed as a simple transformation of the coupling parameter as $\xi \rightarrow \xi (1+w_x)$.  Here we work on the model  
$Q = 3 H \xi (1+w_x) \rho_{x}$ for which the doom factor (\ref{doom}) returns $d = - \xi$, and thus, the stability of this interaction model is ensured for $\xi \geq 0$. One may notice that this interaction function depends on dark energy density as well as pressure like contribution of the dark energy. Here, we focus on the dynamical dark energy equation of state. The main idea is to see how the cosmological parameters are effected in presence of an interaction when dark energy equation of state is dynamical unlike the interaction scenarios with constant equation of state or vacuum interaction. 
To begin with such investigations, we start with the following dark energy parametrization with only one free parameter as  \cite{Gong:2005de}

\begin{eqnarray}
w_x(z)= \frac{w_0}{1+z} \exp \left(\frac{z}{1+z} \right),  \label{eq:ZG-II}
\end{eqnarray}
where $w_0$ is the present value of the dark energy equation of state, i.e., $w_0 = w_x (z =0)$. 
Before closing this section, in Fig. \ref{fig:Q-xi} and \ref{fig:Q-w} we present the qualitative evolution of the present interaction model. In Fig. \ref{fig:Q-xi} we present the qualitative evolution of the interaction function  for several coupling strengths where we analyzed the scenario both for the quintessence (left panel of Fig. \ref{fig:Q-xi}) and phantom dark energy state (right panel of Fig. \ref{fig:Q-xi}) parameters. From the left panel of Fig.  \ref{fig:Q-xi} we see that the interaction function $Q$ remains to be positive, that means, for $w_0 > -1$ regime, the energy flow takes place from CDM to DE. While from the right panel of Fig. \ref{fig:Q-xi} (for $w_0 < -1$), we observe an interesting feature. We see that $w_0 < -1$ enables a sign change in $Q$ in the following way: For high redshifts, $Q > 0$ (energy flow takes place from CDM to DE) while for low redshifts, $Q < 0$ (energy flow thus takes place from DE to CDM). Hence, as a result, for $w_0 < -1$, energy flow changes its direction during the evolution of the universe. 
In Fig. \ref{fig:Q-w} we have analyzed the qualitative nature of $Q$ varying $w_0$ at a fixed value of $\xi$. In the left panel of Fig. \ref{fig:Q-w} we have fixed ($\xi = 0.001$) and varied $w_0$ from the quintessence to phantom regime. It shows that $Q$ remains positive in this case (energy flow from CDM to DE). For the right panel of Fig. \ref{fig:Q-w} we fix $\xi =0.5$ and vary $w_0$. It shows a transition of $Q$ from its negative to positive values and thus indicates towards a change in the direction of energy flow.   

\section{Observational data and the constraints}
\label{sec-data}

In this section we describe the observational data and the fitting mechanism used to constrain the current interacting dark energy models. 

\begin{itemize}

\item We consider the high-$\ell$ temperature and polarization as well as the low-$\ell$ temperature and polarization Cosmic Microwave Background angular power spectra from Planck (Planck TT, TE, EE + lowTEB)  \cite{Adam:2015rua, Aghanim:2015xee}. 

\item The Joint light-curve analysis (JLA) sample from Supernovae Type Ia \cite{Betoule:2014frx}. 

\item Baryon acoustic oscillations (BAO) distance measurements \cite{Beutler:2011hx, Ross:2014qpa,Gil-Marin:2015nqa}. 

\item Hubble parameter measurements from the Cosmic Chronometers (CC)  \cite{Moresco:2016mzx}.

\item Local Hubble constant value yielding $H_0=73.24\pm1.74$ km/s/Mpc at $68 \%$ CL \cite{Riess:2016jrr} from the Hubble Space Telescope (HST).

\item Redshift space distortion data (RSD) \cite{Gil-Marin:2016wya}.

\item Weak lensing (WL) data from the blue galaxy sample compiled from the Canada-France-Hawaii Telescope Lensing Survey \cite{Heymans:2013fya,Asgari:2016xuw}. 

\end{itemize}

In order to extract the observational constraints on the free and derived parameters of the interacting scenario, we perform a fitting analysis using our modified version of the Markov chain Monte Carlo package \texttt{cosmomc} \cite{Lewis:2002ah, Lewis:1999bs} that is equipped with a convergence diagnostic based on the Gelman and Rubin statistic and includes the support for the Planck data release 2015 likelihood Code \cite{Aghanim:2015xee} (see \url{http://cosmologist.info/cosmomc/}) \footnote{Perhaps it should be mentioned here that Planck 2018 results of the cosmological parameters are published \cite{Aghanim:2018eyx}, but the Planck 2018 likelihood is not public yet. Thus, we are unable to use the new likelihood from Planck 2018. However, the use of Planck 2018 data will be worthwhile to update the constraints for the present model as well as for other interacting models.} One can clearly see that the present interacting scenario with one parameter dynamical dark energy equation of state (\ref{eq:ZG-II})  extends its parameters space beyond the six-parameters $\Lambda$CDM model. For convenience, the interacting model with DE assuming the parametrization (\ref{eq:ZG-II}) is labeled as IDE. Thus, one can see that for the spatially flat FLRW universe, the parameters space for IDE is,

\begin{align}
\mathcal{P}_1 \equiv\Bigl\{\Omega_bh^2, \Omega_{c}h^2, 100\theta_{MC}, \tau, w_0, \xi, n_s, log[10^{10}A_S]\Bigr\},
\label{eq:parameter_space1}
\end{align} 
where the parameters $\Omega_bh^2$, $\Omega_{c}h^2$, are respectively the baryon and cold dark matter densities; 
$100 \theta_{MC}$ is the ratio of sound horizon to the angular diameter distance; $\tau$ is the reionization optical depth; $n_s$ is the scalar spectral index; $A_S$ is the amplitude of the primordial scalar power spectrum. The remaining parameters $w_0$ is the free parameter introduced through the parametrization of dark energy equation of state (equation (\ref{eq:ZG-II})). Thus, one can see that the present IDE model is eight dimensional and hence an extended parameters space in compared to the non-interacting $\Lambda$CDM model. The likelihood for this analysis is, $\mathcal{L}\propto e^{-\chi^2/2}$ where $\chi^2 = \sum _{i} \chi^2_i$, and $i$ belongs to the data set $\{$Planck TT, TE, EE $+$ lowTEB, JLA, BAO, CC, HST, RSD, WL$\}$. Thus, one may consider different observational combinations for a detailed analysis of the present interaction model. In what follows, we describe the results of the interacting scenario. The priors on the model parameters for
the statistical analysis are shown in Table \ref{tab:priors}.

\begin{table}
\begin{center}
\begin{tabular}{c|c}
Parameter                    & Prior\\
\hline 
$\Omega_{b} h^2$             & $[0.005,0.1]$\\
$\Omega_{c} h^2$             & $[0.01,0.99]$\\
$\tau$                       & $[0.01,0.8]$\\
$n_s$                        & $[0.5, 1.5]$\\
$\log[10^{10}A_{s}]$         & $[2.4,4]$\\
$100\theta_{MC}$             & $[0.5,10]$\\ 
$\xi$                        & $[0, 1]$\\ 
$w_0$                        & $[-2, 0]$\\
\end{tabular}
\end{center}
\caption{Flat priors on the cosmological parameters for the analysis of the interacting dark energy model. }
\label{tab:priors}
\end{table}

\subsection{Results: IDE}
\label{sec:ide1}

\begin{figure*}
\includegraphics[width=0.65\textwidth]{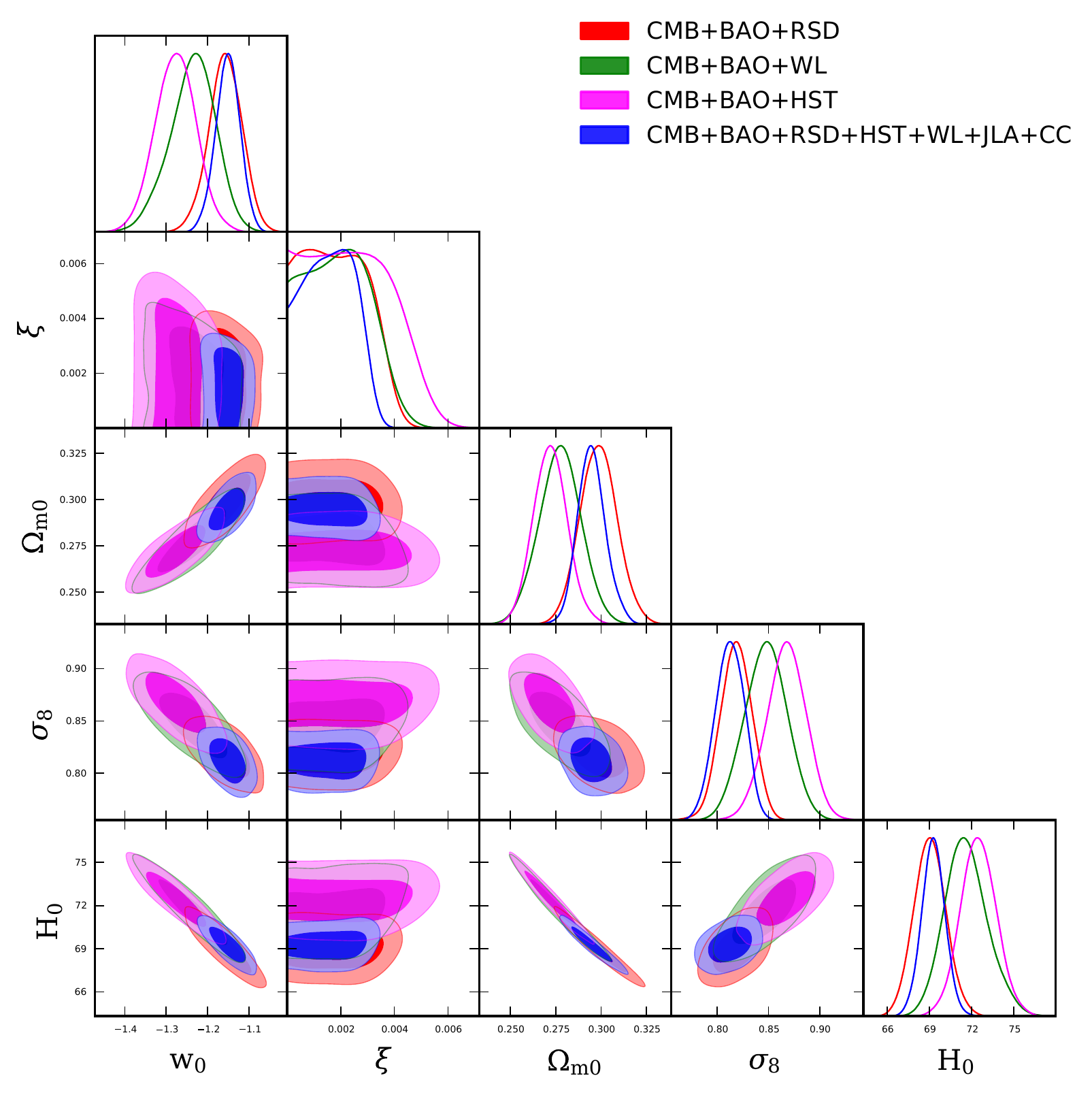}
\caption{The 2D confidence contours of various model parameters as well as their 1D marginalized likelihood functions for the reconstructed interacting  scenarios with the one parameter $w_x(z)$ parametrization given in equation (\ref{eq:ZG-II}). 68.3\% ($1\sigma$) and 95.4\% ($2\sigma$) confidence-level contours and the marginalized likelihood functions are obtained for different combinations of the data sets. Confidence contours clearly show that the coupling parameter $\xi$ is almost uncorrelated with $w_0$ and other parameters. The present value of dark energy equation of state parameter $w_0$ strictly remains in the phantom regime.}
\label{fig-one-parameter-interaction}
\end{figure*}

\begin{figure*}
\includegraphics[width=0.65\textwidth]{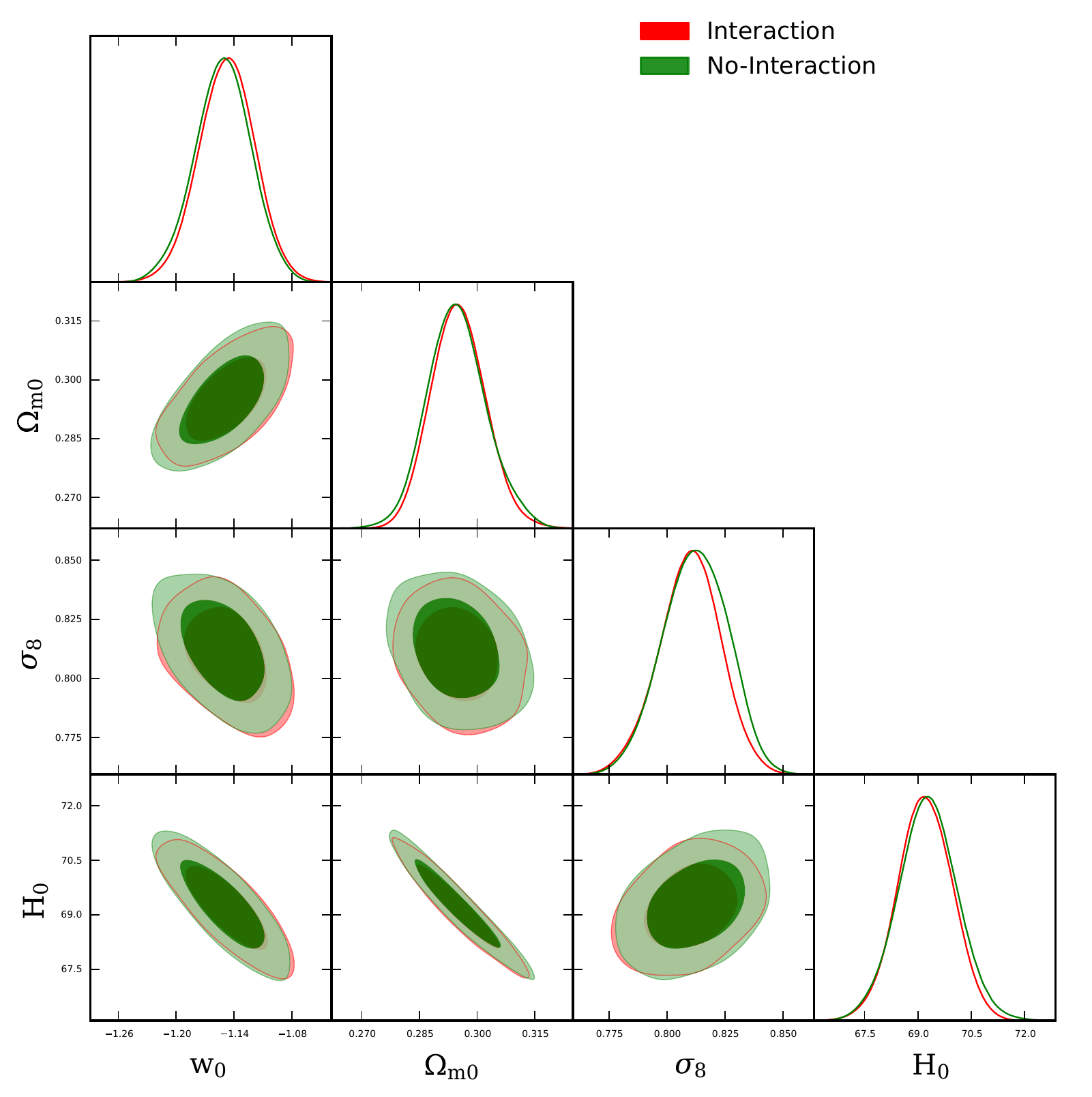}
\caption{The 2D confidence contours of various model parameters as well as their 1D marginalized likelihood functions for the interacting and non-interacting scenarios with the one parameter $w_x(z)$ parametrization given in equation (\ref{eq:ZG-II}). The results of this figure are shown for the combined analysis CMB $+$ BAO $+$ RSD $+$ HST $+$ WL $+$ JLA $+$ CC. One can note that the common parameters of these two cosmological scenarios assume similar values. }
\label{fig-comparison-oneparameter}
\end{figure*}

We summarize the observational constraints on various parameters of IDE in Table~\ref{tab:results-one-parameter-interaction} using different observational datasets. The corresponding confidence level contour plots and also the marginalized likelihood functions are shown in Fig.~\ref{fig-one-parameter-interaction} only for some selected parameters of this interacting model. 

Our analyses show that the value of the coupling parameter $\xi$ is very tiny which eventually makes the interaction function very small compared to the rate of change of energy densities due to the expansion of the universe. 
However, it is interesting to notice that, for some of the combinations, in particular, for CMB $+$ BAO $+$ RSD, CMB $+$ BAO $+$ WL and the full dataset, 
within 68.3\% CL, $\xi \neq 0$ is suggested, while within 95.4\% CL, $\xi = 0$ is definitely allowed. This means that  the non-interacting $w_x$CDM cosmology is recovered in the 95.4\% CL. 
Further, from the estimation of the dark energy equation of state, it is quite clear that the present value of the dark energy equation of state, i.e., $w_0$, is in the phantom regime for all the observational datasets shown in Table \ref{tab:results-one-parameter-interaction}. Even if the CMB data alone allow $w_0$ to be quintessential in the 68.3\% CL, when this dataset is combined with other astronomical probes, $w_0$ is strictly less than `$-1$' at more than $4\sigma$. This result confirms a phantom nature for the DE equation of state, as determined in several works in the literature~\cite{DiValentino:2017iww,DiValentino:2015ola,DiValentino:2016hlg,DiValentino:2017zyq,Mortsell:2018mfj,Yang:2018euj}.

We also observe an interesting feature. From Table \ref{tab:results-one-parameter-interaction}, one may notice that the allowance of the interaction can relieve the tension on $H_0$ as observed from the global \cite{Ade:2015xua} and local measurements \cite{Riess:2016jrr}. Such observation is clearly true for the CMB data alone within $2\sigma$, and thus, it allows us to combine safely HST with the other datasets. Moreover, the combined analysis with CMB $+$ BAO $+$ WL and CMB $+$ BAO $+$ HST alleviates the $H_0$ tension within 68.3\% CL. Further, for the dataset CMB $+$ BAO $+$ RSD, the tension on $H_0$ is released at $2\sigma$. 

In order to examine the effects of the interaction for this particular equation of state in DE (\ref{eq:ZG-II}), we performed the analysis without allowing the interaction.  
In Table \ref{tab:results-one-parameter-without-interaction} we summarize the observational constraints on the free parameters for non-interacting scenario obtained the the same combined analyses as performed for interacting scenario. One can easily see (the Table \ref{tab:results-one-parameter-without-interaction}) that statistically, the cosmological constraints with and without interaction where dark energy has a varying nature given in (\ref{eq:ZG-II}) are really robust. In fact for the interacting model, the coupling strength $\xi$ is found to be uncorrelated with $\Omega_{m0}$, $H_0$, $\sigma_8$ and even with $w_0$, as we can see in Fig. \ref{fig-one-parameter-interaction}. 
In Fig. \ref{fig-comparison-oneparameter}, we display the two dimensional contour plots for various combinations of the model parameters for interacting and non-interacting scenario and also the marginalized likelihood functions for those parameters obtained in the combined analysis with CMB $+$ BAO $+$ RSD $+$ HST $+$ WL $+$ JLA $+$ CC. From Fig. \ref{fig-comparison-oneparameter}, one can easily observe that it is very hard to distinguish the interacting scenario from the non-interacting one. The dark energy equation of state seems to be unaltered even if an interaction is allowed in the dark sectors.

However, it is interesting to notice that without interaction, the CMB data  prefer a phantom dark energy equation of state $w_0$ at more than $1\sigma$, as we can see from Table \ref{tab:results-one-parameter-without-interaction}. Moreover, in this case we find that for all the data sets, $H_0$ has a large shift towards its higher values, which now is fully in agreement with the local value of HST.

\begin{figure*}
k\includegraphics[width=0.65\textwidth]{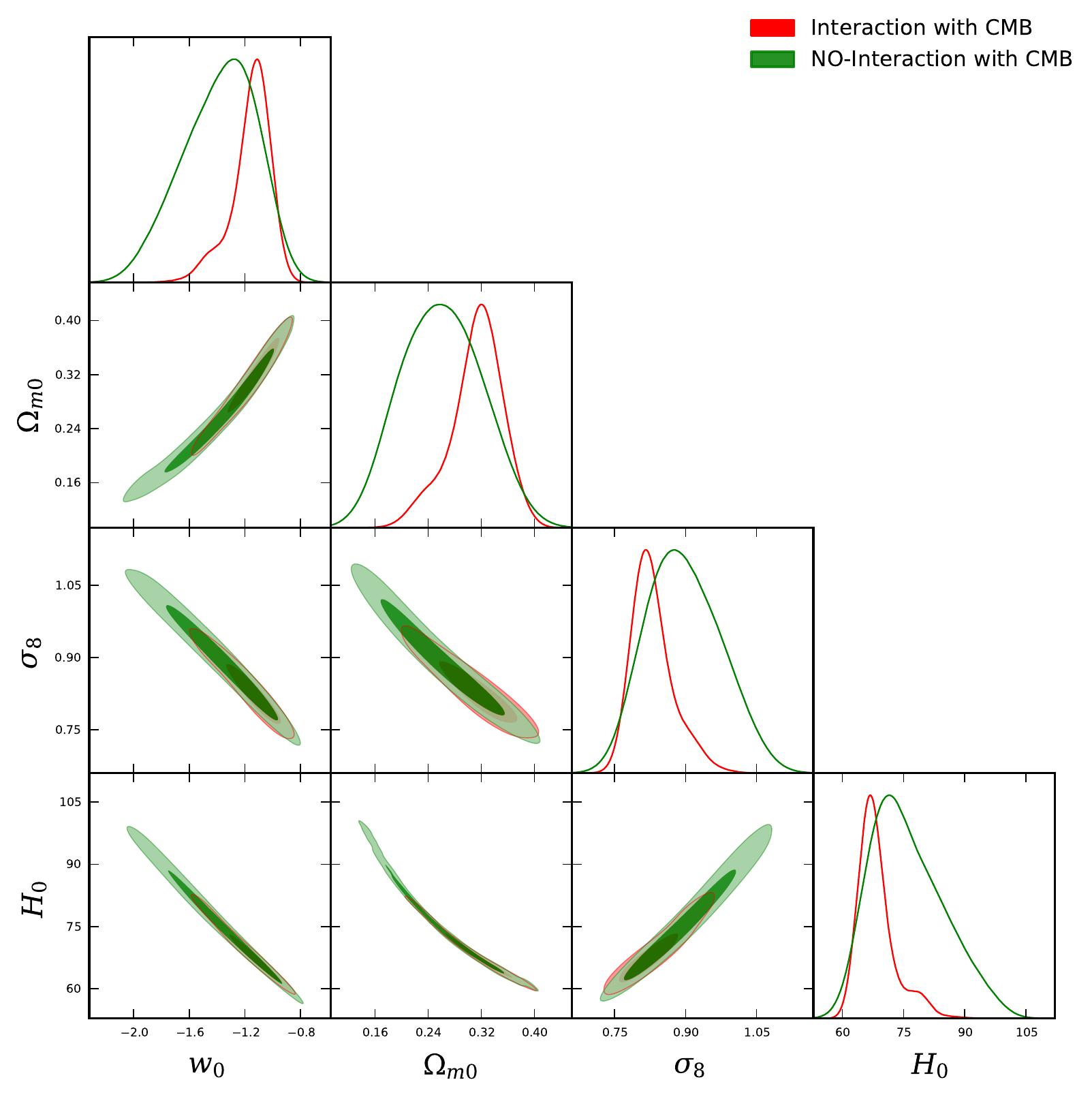}
\caption{The 2D confidence contours of various model parameters as well as their 1D marginalized likelihood functions for the interacting and non-interacting scenarios with the one parameter $w_x(z)$ parametrization given in equation (\ref{eq:ZG-II}). The results of this figure are shown for the CMB data only. One can note that the 2D contours and the 1D marginalized posterior distributions of the parameters for the interacting scenario are shrinked compared to that in the absence of interaction. }
\label{fig-comparison-oneparameter-CMBonly}
\end{figure*}

\begin{figure*}
\includegraphics[width=0.4\textwidth]{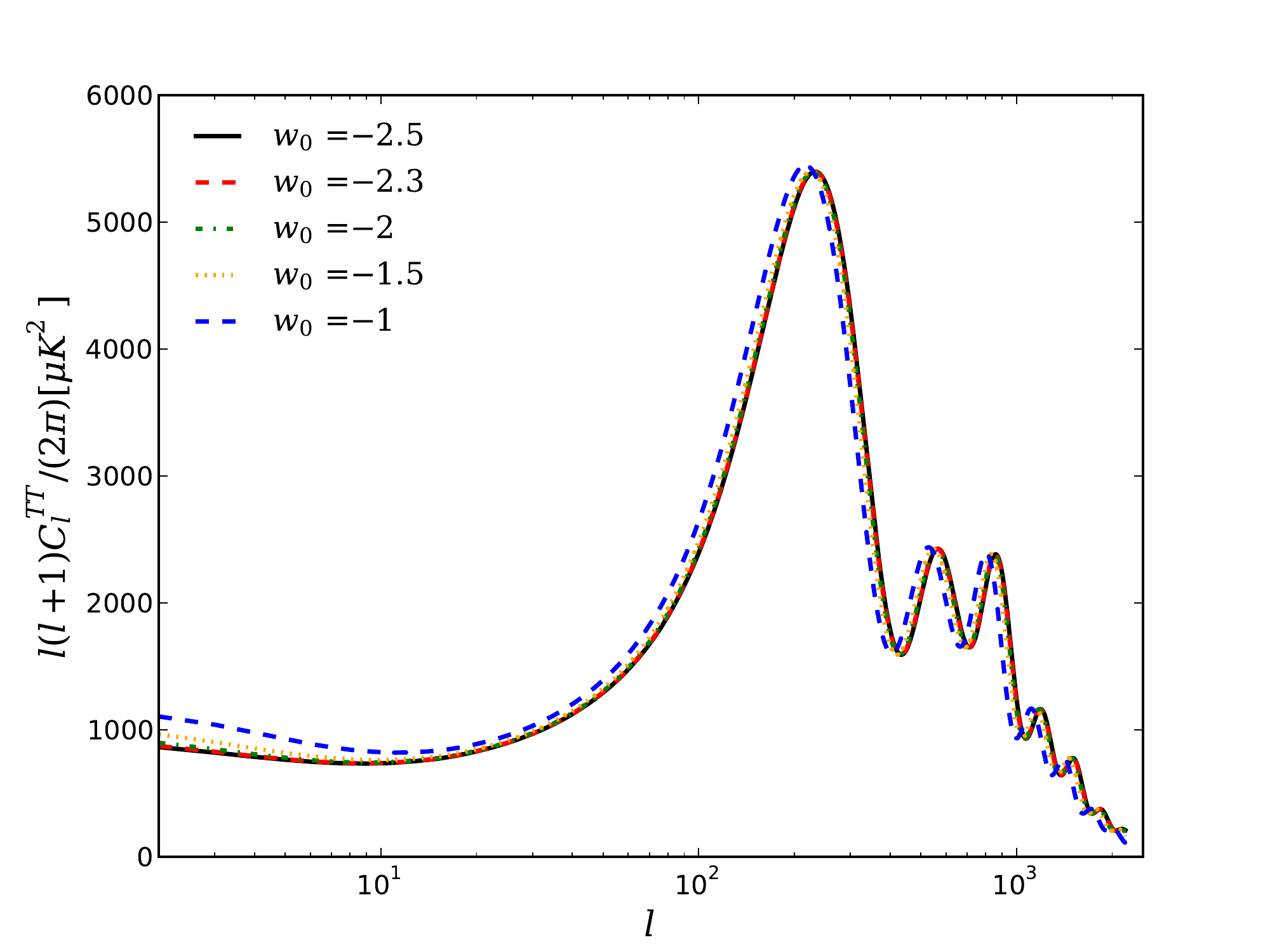}
\includegraphics[width=0.4\textwidth]{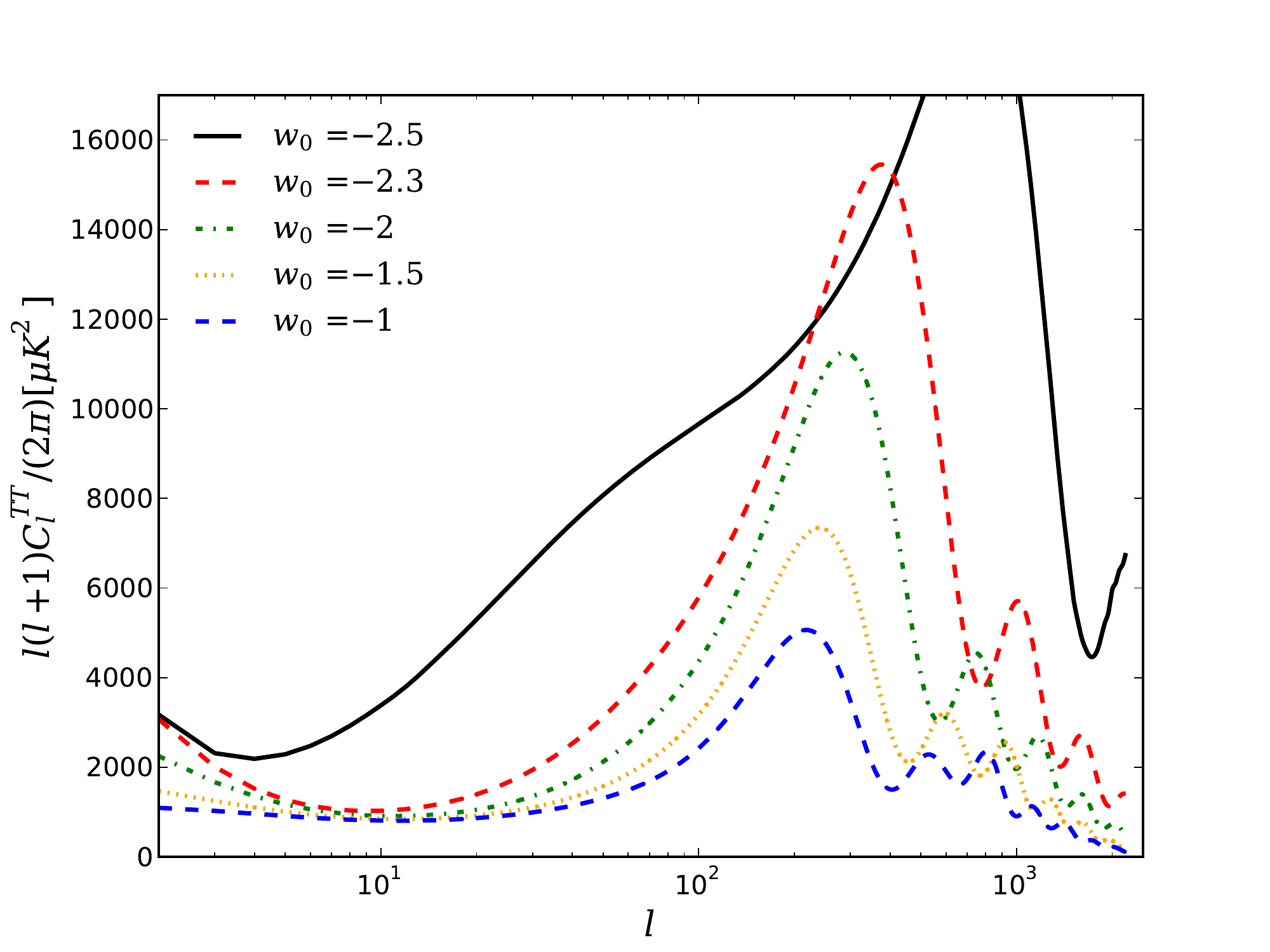}
\caption{The left panel shows the temperature power-spectrum for $\xi=0$ and different values of $w_x (z)$, of equation (\ref{eq:ZG-II}) while the right panel shows the temperature power-spectrum for $\xi=0.5$ and different values of $w_x(z)$. From the left panel one can see that for the non-interaction scenario, no considerable variations are observed even if we increase the magnitude of $w_0$ while from the right panel we see that as long as the interaction is considered into the picture, the variations in the temperature power-spectra become highly sensitive for similar values of $w_0$. }
\label{comp-CMB-TT}
\end{figure*}

Comparing Table~\ref{tab:results-one-parameter-interaction} (interaction) with Table~\ref{tab:results-one-parameter-without-interaction} (no-interaction), we can see that for the CMB only case, the constraints on $w_0$ and the derived parameters become stronger when an interaction is allowed (see also Fig.~\ref{fig-comparison-oneparameter-CMBonly}). This seems to be astonishing as the non-interacting scenario has one less parameter than the interacting scenario. The reason for this feature can be found in Fig.~\ref{comp-CMB-TT}, where it is evident that by introducing the interaction into the cosmological picture, we break the degeneracy for larger negative values of $w_0$, that are no more in agreement with the data. This exact phenomenon is also reflected from the matter power spectrum of the IDE model displayed in Fig. \ref{comp-pk}.

\begingroup

\begin{figure*}
\includegraphics[width=0.4\textwidth]{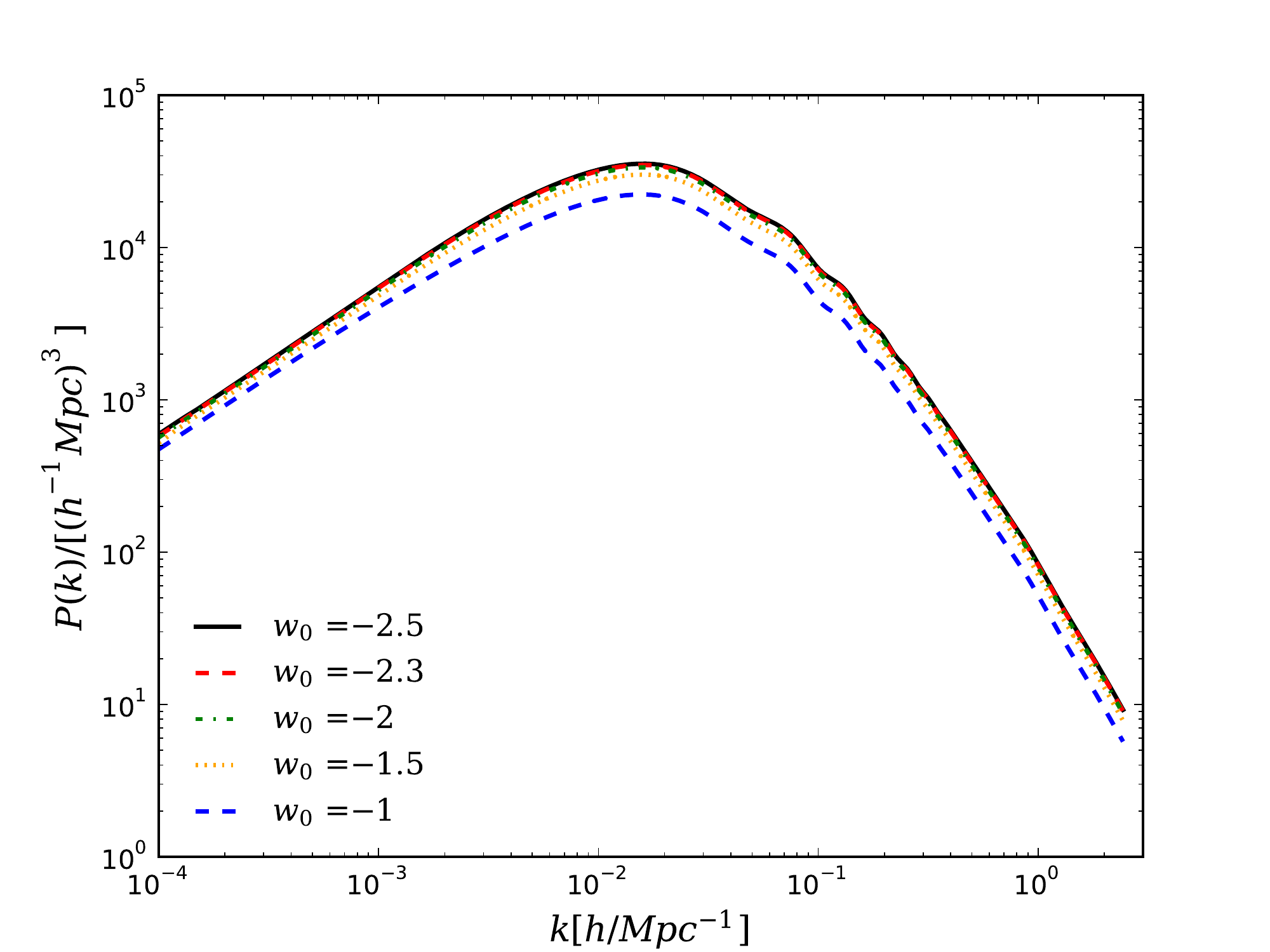}
\includegraphics[width=0.4\textwidth]{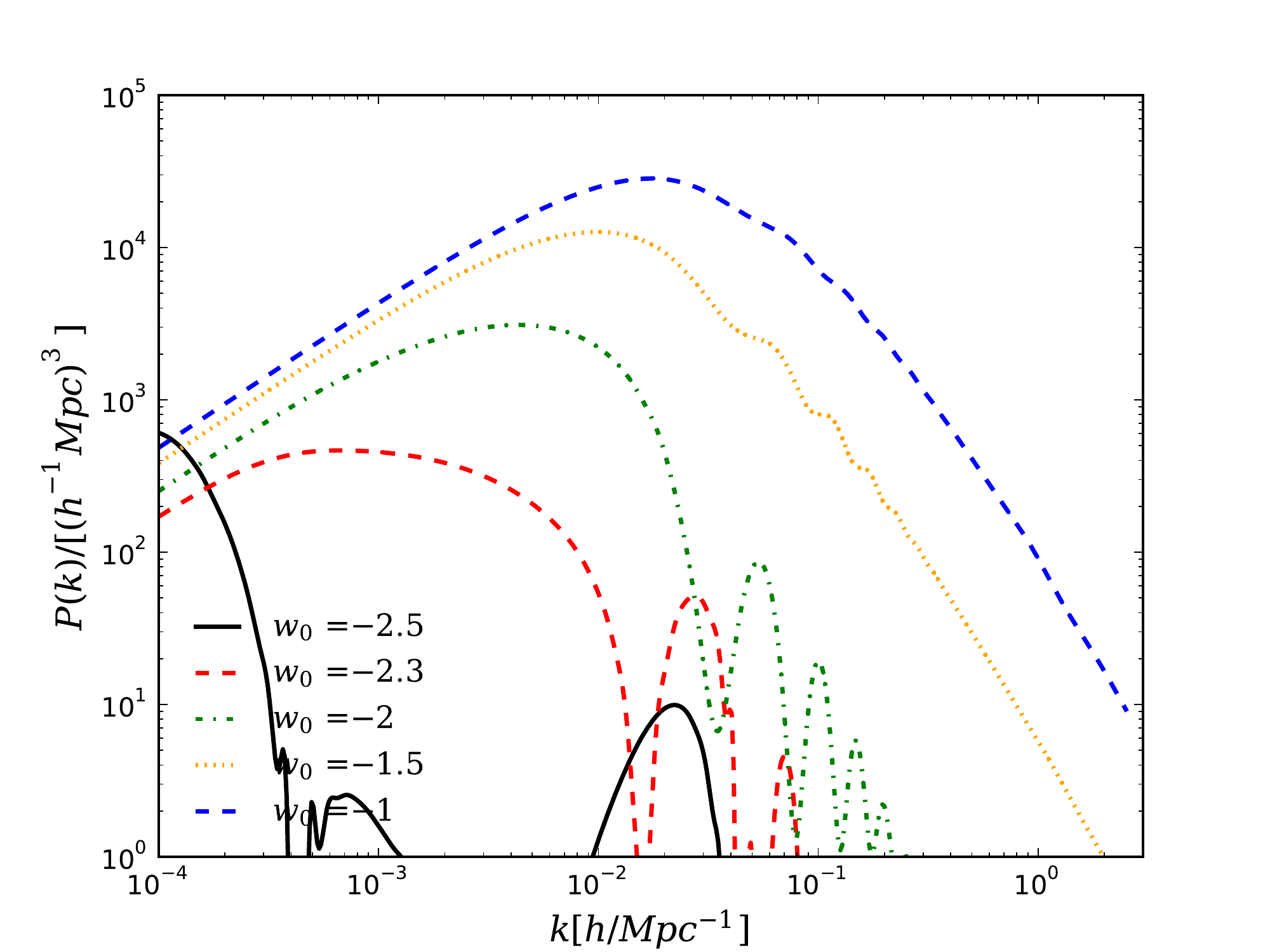}
\caption{The left panel displays the behaviour of the matter power spectrum for $\xi=0$ and different values of $w_x (z)$ of equation (\ref{eq:ZG-II}), while the right panel shows the matter power spectrum with a fixed coupling parameter $\xi = 0.5$ and different values of $w_x(z)$.  Looking at the left panel one can realize that the values of $w_0$ do not affect the matter power spectrum when there is  no interaction, but as long as the interaction is considered, the same values of $w_0$ could change the behaviour of the power spectrum in a vibrant way. }
\label{comp-pk}
\end{figure*}

\begingroup                                                                                                                     
\squeezetable                                                                                                                   
\begin{center}                                                                                                                  
\begin{table*}                                                                                                                   
\begin{tabular}{ccccccccccccccc}                                                                                                            
\hline\hline                                                                                                                    
Parameters & CMB & CMB+BAO+RSD & CMB+BAO+WL & CMB+BAO+HST & $\begin{array}[c]{c}
\text{CMB+BAO+RSD+HST}\\+ \text{WL+JLA+CC} \end{array}$\\ \hline
$\Omega_c h^2$ & $    0.1213_{-    0.0015-    0.0029}^{+    0.0014+    0.0030}$ & $    0.1194_{-    0.0012-    0.0023}^{+    0.0012+    0.0024}$ &  $    0.1192_{-    0.0012-    0.0024}^{+    0.0013+    0.0022}$ & $    0.1199_{-    0.0012-    0.0025}^{+    0.0012+    0.0024}$ & $    0.1185_{-    0.0011-    0.0022}^{+    0.0011+    0.0022}$\\

$\Omega_b h^2$ & $ 0.02208_{-    0.00017-    0.00032}^{+    0.00016+    0.00033}$ & $    0.02219_{-    0.00018-    0.00028}^{+    0.00014+    0.00030}$ & $0.02223_{-    0.00015-    0.00026}^{+    0.00013+    0.00029}$ & $ 0.02218_{-    0.00014-    0.00029}^{+    0.00015+    0.00028}$ & $ 0.02227_{-    0.00014-    0.00028}^{+    0.00014+    0.00028}$ \\

$100\theta_{MC}$ & $ 1.04019_{-    0.00034-    0.00071}^{+    0.00038+    0.00066}$ & $    1.04042_{-    0.00031-    0.00064}^{+    0.00033+    0.00065}$ & $1.04050_{-    0.00031-    0.00060}^{+    0.00030+    0.00062}$ & $ 1.04043_{-    0.00032-    0.00063}^{+    0.00033+    0.00060}$ & $ 1.04053_{-    0.00031-    0.00064}^{+    0.00034+    0.00058}$\\

$\tau$ & $    0.085_{-    0.019-    0.032}^{+    0.017+    0.034}$ & $    0.079_{-    0.019-    0.034}^{+    0.018+    0.036}$ & $    0.082_{-    0.018-    0.036}^{+    0.019+    0.035}$ & $    0.086_{-    0.017-    0.033}^{+    0.017+    0.034}$ & $    0.078_{-    0.018-    0.033}^{+    0.018+    0.034}$\\

$n_s$ & $    0.9710_{-    0.0045-    0.0085}^{+    0.0041+    0.0091}$ & $    0.9750_{-    0.0042-    0.0089}^{+    0.0042+    0.0082}$ & $    0.9757_{-    0.0042-    0.0079}^{+    0.0042+    0.0081}$ & $    0.9738_{-    0.0041-    0.0078}^{+    0.0041+    0.0083}$ &  $    0.9771_{-    0.0040-    0.0079}^{+    0.0040+    0.0081}$\\

${\rm{ln}}(10^{10} A_s)$ & $    3.114_{-    0.035-    0.063}^{+    0.035+    0.066}$ & $    3.097_{-    0.037-    0.067}^{+    0.034+    0.069}$ & $    3.105_{-    0.034-    0.072}^{+    0.035+    0.068}$ & $    3.116_{-    0.033-    0.065}^{+    0.035+    0.065}$ & $    3.093_{-    0.034-    0.065}^{+    0.034+    0.066}$\\

$w_0$ & $   -1.16_{-    0.09-    0.35}^{+    0.17+    0.26}$ & $   -1.157_{-    0.039-    0.083}^{+    0.040+    0.076}$ & $   -1.24_{-    0.05-    0.12}^{+    0.06+    0.10}$ & $   -1.277_{-    0.049-    0.097}^{+    0.049+    0.097}$ & $   -1.151_{-    0.029-    0.061}^{+    0.028+    0.057}$ \\

$\xi$ & $    <0.0028\, <0.0046 $ & $    0.0019_{-    0.0017}^{+    0.0009}\,<0.0037$ & $    0.0020_{-    0.0015}^{+    0.0011}\,<0.0038$ & $    <0.0032\,<0.0048$ & $    0.0016_{-    0.0009}^{+    0.0011}\,<0.0030$\\

$\Omega_{m0}$ & $    0.3130_{-    0.0306-    0.0938}^{+    0.0463+    0.0765}$ & $    0.2986_{-    0.0101-    0.0199}^{+    0.0099+    0.0205}$ & $    0.2774_{-    0.0114-    0.0227}^{+    0.0112+    0.0221}$ & $    0.2720_{-    0.0094-    0.0182}^{+    0.0092+    0.0194}$ & $    0.2948_{-    0.0084-    0.0141}^{+    0.0071+    0.0164}$\\

$\sigma_8$ & $    0.829_{-    0.054-    0.080}^{+    0.029+    0.10}$ & $    0.819_{-    0.015-    0.030}^{+    0.015+    0.029}$ & $    0.848_{-    0.021-    0.040}^{+    0.021+    0.039}$ & $    0.868_{-    0.019-    0.039}^{+    0.019+    0.036}$ & $    0.812_{-    0.014-    0.028}^{+    0.015+    0.027}$\\

$H_0$ & $   68.3_{-    5.4-    8.1}^{+    2.4+   12}$ & $   69.0_{-    1.1-    2.2}^{+    1.1+    2.3}$ & $   71.6_{-    1.7-    2.9}^{+    1.4+    3.2}$ & $   72.5_{-    1.2-    2.5}^{+    1.3+    2.5}$ & $   69.28_{-    0.82-    1.7}^{+    0.82+    1.6}$\\
\hline\hline                                                                                                                    
\end{tabular}                                                                                                                   
\caption{68.3\% ($1\sigma$) and 95.4\% ($2\sigma$) constraints on the model parameters for the interacting dark matter-dark energy scenario where the dark energy equation of state has a single free parameter shown in equation (\ref{eq:ZG-II}). }
\label{tab:results-one-parameter-interaction}                                                                                                   
\end{table*}                                                                                                                     
\end{center}                                                                                                                    
\endgroup                                                                                                                       

\begingroup                                                                                                                     
\squeezetable                                                                                                                   
\begin{center}                                                                                                                  
\begin{table*}                                                                                                                   
\begin{tabular}{cccccccccccccccccccc}                                                                                                            
\hline\hline                                                                                                                    
Parameters & CMB & CMB+BAO+RSD & CMB+BAO+WL & CMB+BAO+HST & $\begin{array}[c]{c}
\text{CMB+BAO+RSD+HST}\\+ \text{WL+JLA+CC} \end{array}$ \\ \hline

$\Omega_c h^2$ & $    0.1205_{-    0.0015-    0.0032}^{+    0.0016+    0.0032}$ & $    0.1194_{-    0.0011-    0.0023}^{+    0.0011+    0.0022}$ & $    0.1191_{-    0.0012-    0.0022}^{+    0.0012+    0.0023}$ & $    0.1196_{-    0.0012-    0.0022}^{+    0.0011+    0.0023}$ & $    0.1184_{-    0.0011-    0.0022}^{+    0.0011+    0.0023}$\\

$\Omega_b h^2$ & $0.02215_{-    0.00017-    0.00034}^{+    0.00017+    0.00035}$ & $    0.02220_{-    0.00016-    0.00028}^{+    0.00015+    0.00029}$ & $ 0.02222_{-    0.00014-    0.00027}^{+    0.00015+    0.00027}$ & $ 0.02219_{-    0.00014-    0.00027}^{+    0.00014+    0.00028}$ & $ 0.02227_{-    0.00016-    0.00028}^{+    0.00014+    0.00029}$ \\

$100\theta_{MC}$ & $ 1.04034_{-    0.00034-    0.00071}^{+    0.00035+    0.00070}$ & $    1.04045_{-    0.00032-    0.00061}^{+    0.00032+    0.00063}$ & $1.04047_{-    0.00031-    0.00061}^{+    0.00030+    0.00059}$ & $ 1.04045_{-    0.00032-    0.00061}^{+    0.00032+    0.00063}$ &  $ 1.04057_{-    0.00036-    0.00062}^{+    0.00029+    0.00067}$\\

$\tau$ & $    0.082_{-    0.018-    0.035}^{+    0.019+    0.035}$ & $    0.080_{-    0.018-    0.035}^{+    0.019+    0.035}$ & $    0.085_{-    0.018-    0.035}^{+    0.018+    0.035}$ & $    0.086_{-    0.017-    0.036}^{+    0.017+    0.035}$ & $    0.078_{-    0.020-    0.035}^{+    0.018+    0.035}$\\

$n_s$ & $    0.9721_{-    0.0047-    0.0087}^{+    0.0045+    0.0089}$ & $    0.9751_{-    0.0042-    0.0084}^{+    0.0042+    0.0081}$ & $    0.9757_{-    0.0042-    0.0082}^{+    0.0042+    0.0079}$ & $    0.9745_{-    0.0046-    0.0079}^{+    0.0042+    0.0081}$ & $    0.9776_{-    0.0039-    0.0077}^{+    0.0040+    0.0078}$ \\

${\rm{ln}}(10^{10} A_s)$ & $    3.108_{-    0.035-    0.070}^{+    0.035+    0.067}$ & $    3.099_{-    0.036-    0.070}^{+    0.036+    0.070}$ & $    3.110_{-    0.035-    0.068}^{+    0.034+    0.067}$ & $    3.113_{-    0.033-    0.069}^{+    0.034+    0.066}$ & $    3.093_{-    0.039-    0.065}^{+    0.035+    0.069}$\\

$w_0$ & $   -1.38_{-    0.23-    0.55}^{+    0.33+    0.49}$ & $   -1.153_{-    0.037-    0.074}^{+    0.038+    0.071}$ & $   -1.234_{-    0.054-    0.12}^{+    0.063+    0.11}$ & $   -1.263_{-    0.047-    0.097}^{+    0.053+    0.10}$ & $   -1.15_{-    0.030-    0.058}^{+    0.029+    0.056}$\\

$\Omega_{m0}$ & $    0.260_{-    0.076-    0.12}^{+    0.057+    0.12}$ & $    0.2998_{-    0.0092-    0.019}^{+    0.0093+    0.018}$ & $    0.278_{-    0.012-    0.023}^{+    0.012+    0.023}$ & $    0.2738_{-    0.0097-    0.018}^{+    0.0092+    0.020}$ &  $    0.2952_{-    0.0077-    0.013}^{+    0.0070+    0.014}$\\

$\sigma_8$ & $    0.896_{-    0.096-    0.15}^{+    0.074+    0.16}$ & $    0.818_{-    0.016-    0.032}^{+    0.015+    0.032}$ & $    0.849_{-    0.023-    0.040}^{+    0.021+    0.043}$ & $    0.862_{-    0.020-    0.039}^{+    0.020+    0.040}$ & $    0.810_{-    0.013-    0.028}^{+    0.014+    0.026}$ \\

$H_0$ & $   76_{-   12-   16}^{+7.0 + 19}$ & $   68.9_{-    1.0-    1.9}^{+    1.0+    2.1}$ & $   71.5_{-    1.7-    3.0}^{+    1.4+    3.3}$ & $   72.1_{-    1.3-    2.5}^{+    1.3+    2.5}$ & $   69.20_{-    0.73-    1.6}^{+    0.76+    1.5}$\\

\hline\hline                                                                                                                    
\end{tabular}                                                                                                                   
\caption{68.3\% ($1\sigma$) and 95.4\% ($2\sigma$) constraints on the model parameters for the non-interacting dark matter-dark energy scenario (i.e., $\xi =0$) in which the dark energy state parameter follows equation (\ref{eq:ZG-II}). }
\label{tab:results-one-parameter-without-interaction}                                                                                                   
\end{table*}                                                                                                                     
\end{center}                                                                                                                    
\endgroup

\begin{figure*}
\includegraphics[width=0.4\textwidth]{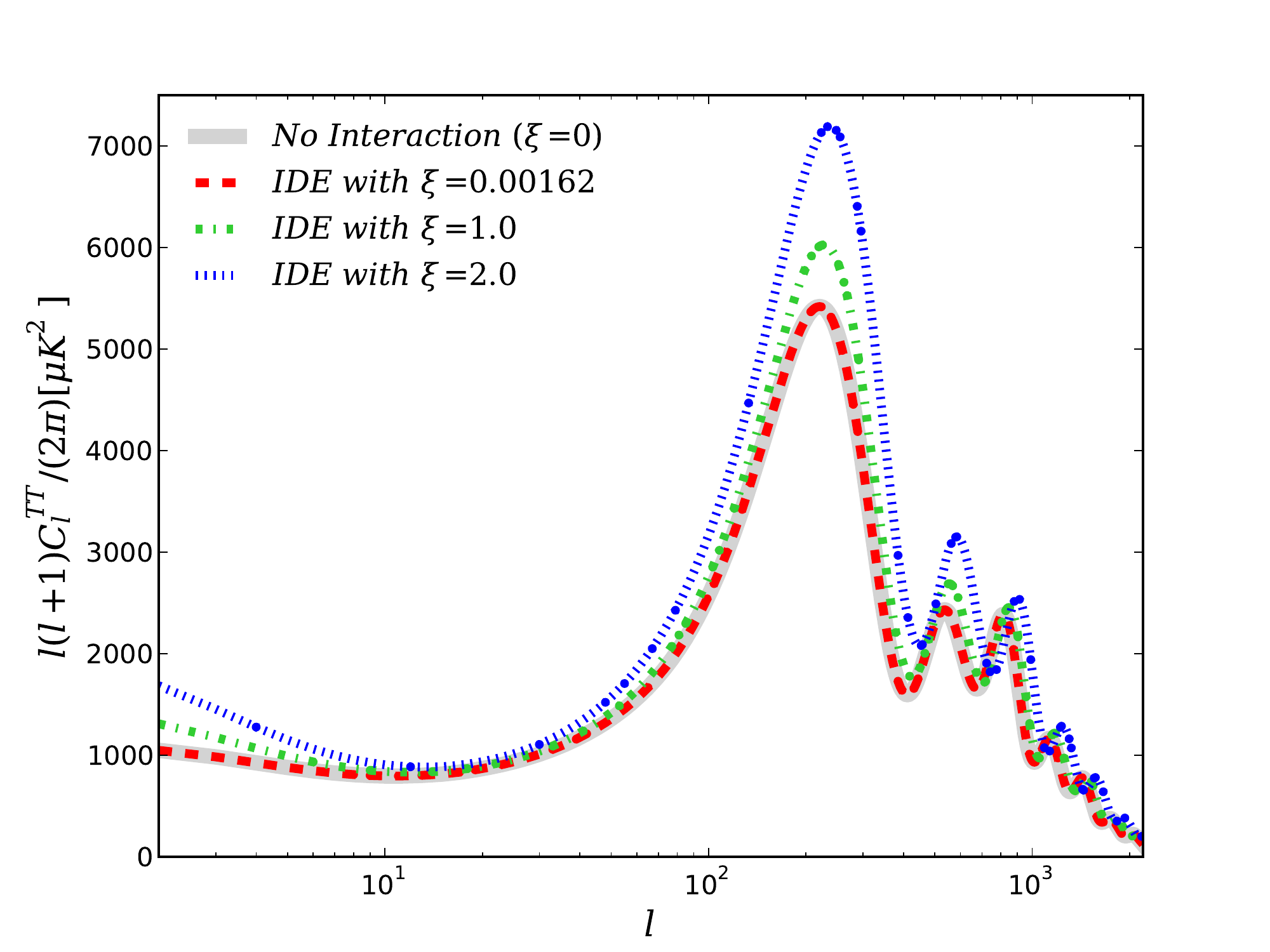}
\includegraphics[width=0.4\textwidth]{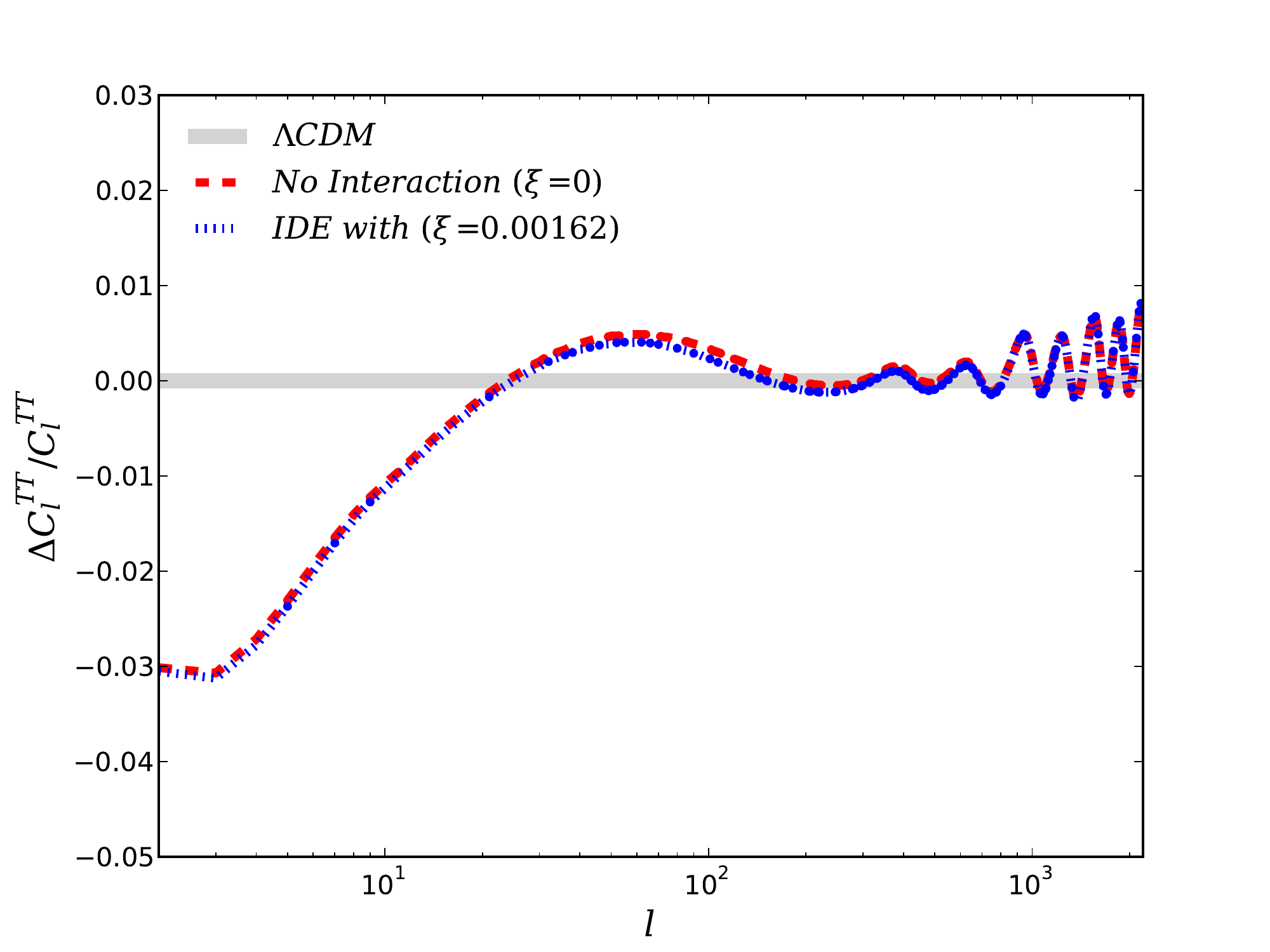}
\caption{The left panel shows the power-spectrum of temperature anisotropy for the non-interacting ($\xi = 0$) and interacting dark energy model (for different values of the coupling parameter $\xi$) where dark energy has a dynamical state parameter given in eqn. (\ref{eq:ZG-II}). The right panel shows the corresponding residuals for the model for both interacting and non-interacting scenario with respect to the base $\Lambda$CDM model. For both left and right panels, the parameters, such as $w_0$ and others are fixed according to their mean values obtained from the combined analysis CMB $+$ BAO $+$ RSD $+$ HST $+$ WL $+$ JLA $+$ CC.  From the left panel one can clearly see that as $\xi$ increases, the height of the first acoustic peak in the power spectrum increases compared to the non-interaction scenario. }
\label{fig-cmb-power}
\end{figure*}

\begin{figure*}
\includegraphics[width=0.4\textwidth]{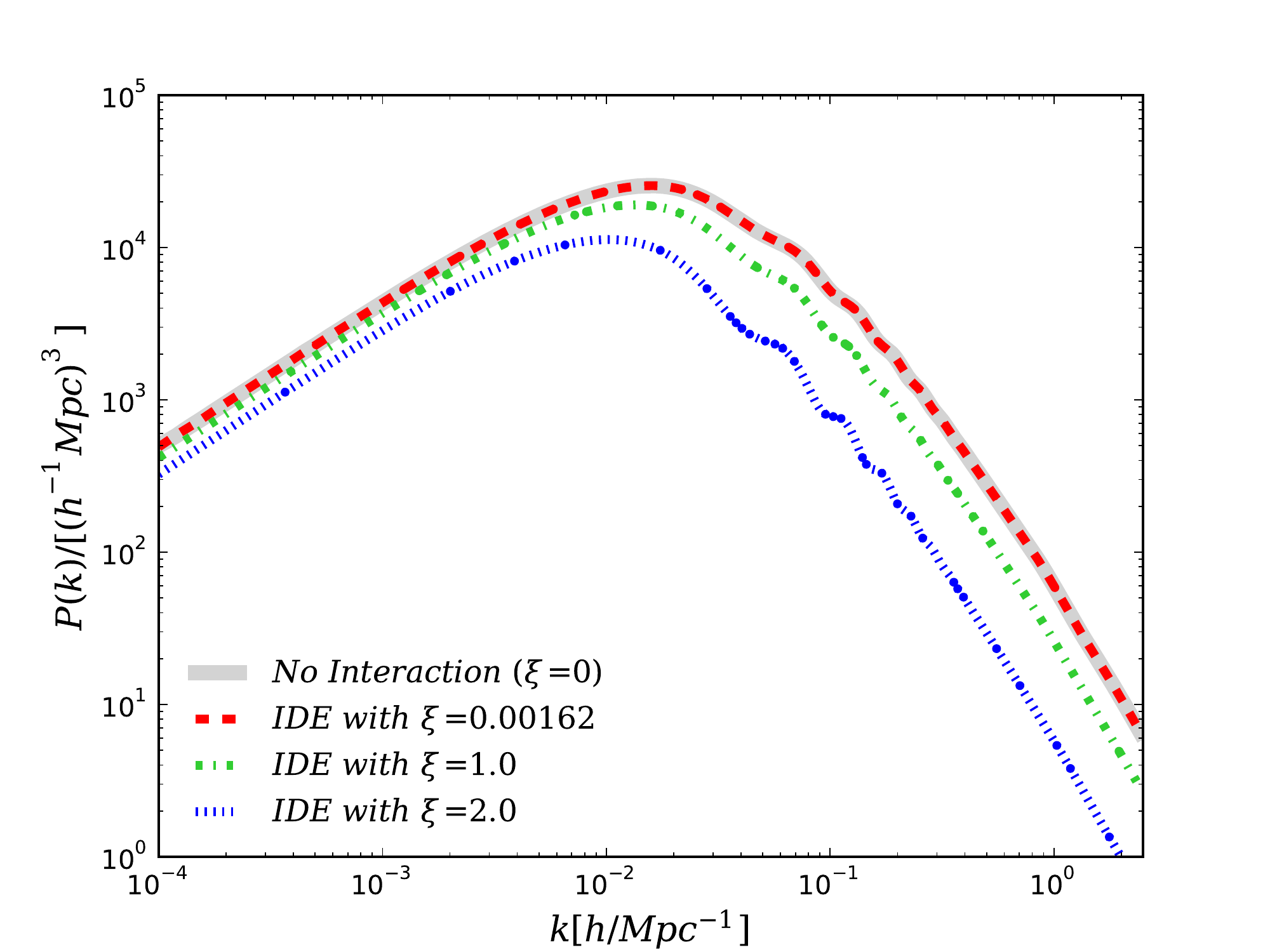}
\includegraphics[width=0.4\textwidth]{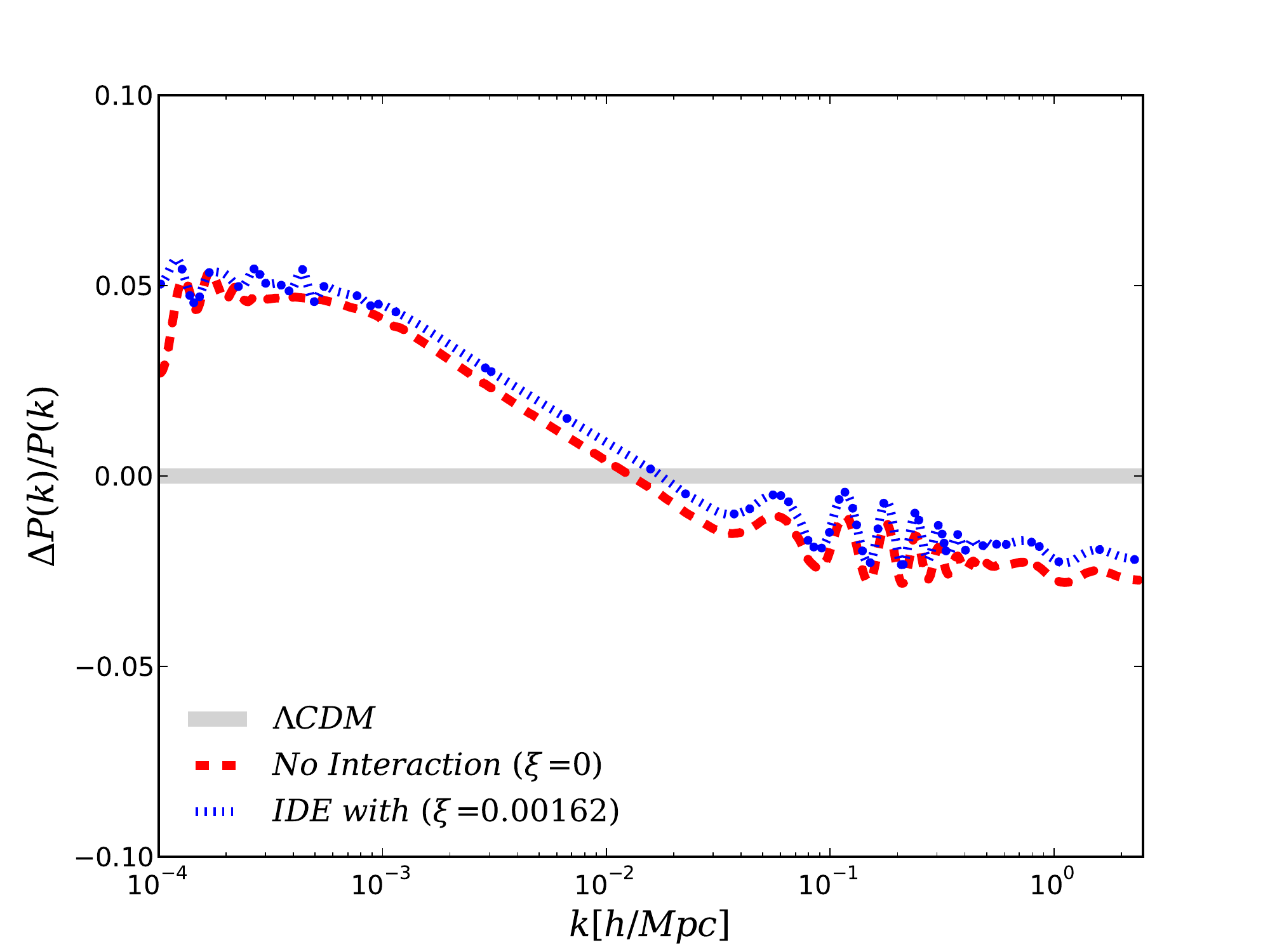}
\caption{The left panel shows the power-spectrum of the matter density contrast for the non-interacting ($\xi = 0$) and interacting dark energy model (for different values of the coupling parameter $\xi$) where dark energy has a dynamical state parameter given in eqn. (\ref{eq:ZG-II}). The right panel shows the corresponding residuals for both interacting and non-interacting scenarios with respect to the base $\Lambda$CDM model. For both left and right panels, the parameters, such as $w_0$ and others are fixed according to their mean values obtained from the combined analysis CMB $+$ BAO $+$ RSD $+$ HST $+$ WL $+$ JLA $+$ CC. From the left panel one can see that the matter power spectrum gets suppressed as the strength of the interaction increases.}
\label{fig-matter-power}
\end{figure*}
\begingroup                                                                                                                     
\begin{center}                                                                                                                  
\begin{table*}                                                                                                                
\begin{tabular}{ccc}                                                                                                            
\hline\hline                                                                                                                    
$\ln B_{ij}$ & ~~~~~~~Strength of evidence for model ${M}_i$ \\ \hline
$0 \leq \ln B_{ij} < 1$ & Weak \\
$1 \leq \ln B_{ij} < 3$ & Definite/Positive \\
$3 \leq \ln B_{ij} < 5$ & Strong \\
$\ln B_{ij} \geq 5$ & Very strong \\
\hline\hline                                                                                                                    
\end{tabular}                                                                                                                   
\caption{Interpretation of the revised Jeffreys scale by Kass and Raftery \cite{Kass:1995loi} used in this work.} \label{tab:jeffreys}                                                                                                   
\end{table*}                                                                                                                     
\end{center}                                                                                                                    
\endgroup

\begingroup                                                                                                                     
\begin{center}                                                                                                                  
\begin{table*}                                                                                                                
\begin{tabular}{cccc}                                                                                                            
\hline\hline                                                                                                                    
Dataset & Model & $\ln B_{ij}$ & ~~~~~~~Strength of evidence for model $\Lambda$CDM \\ \hline

CMB & IDE & -4.3 & Strong \\
CBR & IDE & -4.8 & Strong \\
CBW & IDE & -6.7 & Very Strong\\
CBH & IDE & -8.5 & Very Strong\\
CBRHWJC & IDE & -3.9 & Strong \\

\hline\hline                                                                                                                    
\end{tabular}                                                                                                                   
\caption{Values of $\ln {\cal B}$ and the strength of the evidence for the IDE model against the $\Lambda$CDM, as obtained in our analysis for different dataset combinations. The negative sign actually indicates that the $\Lambda$CDM is preferred over the IDE model. We note that here, CBR = CMB $+$ BAO $+$ RSD, CBW = CMB $+$ BAO $+$ WL, CBH = CMB $+$ BAO $+$ HST, and CBRHWJC = CMB $+$ BAO $+$ RSD $+$ HST $+$ WL $+$ JLA $+$ CC.  }\label{tab:bayesian}                                                                                                   
\end{table*}                                                                                                                     
\end{center}                                                                                                                    
\endgroup

\subsection{IDE at large scales: CMB and matter power spectra}
\label{sec-cmb+matter}

Let us now discuss the behaviour of the IDE model at large scales of the universe and also measure the deviation from its corresponding non-interacting scenario. In order to do so, in the left panel of Fig.~\ref{fig-cmb-power}, we show the temperature anisotropy in the CMB spectra for the IDE scenario (using different values of the coupling parameter $\xi$) and the non-interacting scenario (corresponds to $\xi = 0$). The essential parameters such as $w_0$ and others, used for drawing the plot are fixed according to their mean values, obtained from the combined analysis CMB $+$ BAO $+$ RSD $+$ HST $+$ WL $+$ JLA $+$ CC. In the right panel of Fig.~\ref{fig-cmb-power} we show the corresponding residual with respect to the base $\Lambda$CDM model where we also include the case for non-interacting scenario. From the left panel of Fig.~\ref{fig-cmb-power}, one can notice that the height of the first acoustic peak in the CMB power-spectrum changes as the coupling strength increases. The corresponding residuals for the model (right panel of Fig.~\ref{fig-cmb-power}) behave in a  similar way for higher multipoles  (around $l\sim 10^3$), but at lower multipoles, in the cosmic variance limited region around $l\sim 10$, the one parameter $w_x$ model (\ref{eq:ZG-II}) sharply deviates from the base $\Lambda$CDM for both interacting and non-interacting cases. 

In the left panel of Fig.~\ref{fig-matter-power}, we show the power spectra of the matter density contrast for the interacting model for different values of the coupling parameter $\xi$ where $\xi = 0$ refers to the corresponding non-interacting model. In the right panel of  Fig.~\ref{fig-matter-power},  we display the corresponding residual with respect to the base $\Lambda$CDM model scaled by the corresponding $\Lambda$CDM values. We note that the essential parameters such as $w_0$ and others, used for drawing the plot are fixed according to their mean values, obtained from the combined analysis CMB $+$ BAO $+$ RSD $+$ HST $+$ WL $+$ JLA $+$ CC.  From the left panel of  Fig.~\ref{fig-matter-power}, one can see that the matter power spectrum for this one parameter dark energy state parameter (\ref{eq:ZG-II}) gets suppressed with the increase of the interaction strength quantified by $\xi$.

\subsection{Bayesian Evidence}
\label{sec-bayesian}

In this section, statistical comparison of models has been discussed by calculating  the evidence of the present interacting dark energy model with respect to the reference $\Lambda$CDM model following the Bayesian analysis.   

In Bayesian analysis, the posterior probability distribution of a model parameter $\theta$ is defined based on a given data set $x$, used to test the model $M$, and any prior information. The Bayes theorem states the posterior probability of the parameter $\theta$ as,
\begin{eqnarray}\label{BE}
p(\theta|x, M) = \frac{p(x|\theta, M)\,\pi(\theta|M)}{p(x|M)},
\end{eqnarray}
where the quantity $p(x|\theta, M)$ represents the likelihood function  which depends on the model parameters $\theta$ with the fixed data set and $\pi(\theta|M)$ is the prior information. The quantity $p(x|M)$ in the denominator of the right hand side of eq.~(\ref{BE}) 
is known as the Bayesian evidence which is actually the integral 
over the unnormalised posterior $\tilde{p} (\theta|x, M) \equiv p(x|\theta,M)\,\pi(\theta|M)$. It expressed as 

\begin{eqnarray}\label{sp-be01}
E \equiv p(x|M) = \int d\theta\, p(x|\theta,M)\,\pi(\theta|M). 
\end{eqnarray}
This is also referred to as the global likelihood.  Now, between any two models, $M_i$ and $M_j$ where 
$M_i$ is the model under consideration and $M_j$ is the reference model (here the $\Lambda$CDM model), the posterior probability is given by the product of the ratio of the model priors and
the  ratio  of  Evidences 

\begin{eqnarray}
\frac{p(M_i|x)}{p(M_j|x)} = \frac{\pi(M_i)}{\pi(M_j)}\,\frac{p(x| M_i)}{p(x|M_j)} = \frac{\pi(M_i)}{\pi(M_j)}\, B_{ij}.
\end{eqnarray}
where $B_{ij} = \frac{p(x| M_i)}{p(x|M_j)}$, is the Bayes factor of the considered model $M_i$ compared to the reference model $M_j$. This factor reports how the observational data support the model $M_i$ over $M_j$.  We classify the model comparison as follows: if $B_{ij} > 1 $, then the data support the model $M_i$ more strongly compared to the model $M_j$. Now, depending on different values of the Bayes factor $B_{ij}$ (sometimes one calculates the values of $\ln B_{ij}$), we compare the models. This quantification is generally done accepting the revised Jeffreys scale by Kass and Raftery~\cite{Kass:1995loi} displayed in Table~\ref{tab:jeffreys}.

The Bayesian evidence is computed using the MCMC chains for the statistical analysis of the model.  We refer to the original works \cite{Heavens:2017hkr,Heavens:2017afc} for a detailed implementation of the code \texttt{MCEvidence}\footnote{This code is publicly available at \href{https://github.com/yabebalFantaye/MCEvidence}{github.com/yabebalFantaye/MCEvidence}.}.

In Table \ref{tab:bayesian} we present the $\ln B_{ij}$ values of the IDE model with respect to the base $\Lambda$CDM model. The negative values of $\ln B_{ij}$ indicate the preference of the $\Lambda$CDM model over the IDE model. From the numerical values of $\ln B_{ij}$ and using the Jeffreys scale (Table \ref{tab:jeffreys}), one may clearly conclude that the base $\Lambda$CDM is strongly favored over the present IDE model.

\section{Conclusions and Final remarks}
\label{sec-conclu}

Cosmological models where the main two dark fluids of the universe namely dark matter and dark energy interact with each other is the main concern of this work. Theoretically, an interaction in the dark sector is able to provide some explanations for the cosmological constant problem \cite{Wetterich-ide1} and cosmic coincidence problem \cite{Amendola-ide1,Amendola-ide2,Pavon:2005yx,delCampo:2008sr,delCampo:2008jx}. From the observational point of view, the possibility of an interaction in the dark sector is indicated by a series of latest astronomical data \cite{Salvatelli:2014zta, Nunes:2016dlj,Kumar:2016zpg, vandeBruck:2016hpz, Yang:2017yme, Kumar:2017dnp, DiValentino:2017iww}. Additionally, an interaction in the dark sector is also able to reconcile the $H_0$ tension \cite{DiValentino:2017iww, Kumar:2016zpg,Yang:2017zjs, Bhattacharyya:2018fwb}. Thus, the cosmological models allowing an interaction in the dark sector, are gaining significant attention at current time. In this work we investigate an interacting dark energy scenario where the dark energy has a time varying equation of state parameter (equation (\ref{eq:ZG-II})). We note that the interacting scenarios where dark energy has a dynamical equation of state are clearly the most general ones compared to the interacting scenarios with $w_x =$ constant.  However, the theory of interaction is not smooth enough since the incorporation of an interaction is equally responsible in affecting the other observables. One important issue, related to the interacting dark energy, is the large scale stability of the model which depends on the choice of the interaction function. Most of the interacting dark energy models, present in the literature, suffers from the singularity at $w_x=-1$. But the problem can be alleviated with some particular choices of the interaction function \cite{Yang:2017zjs, Yang:2017ccc}. Hence, in this work, the interaction function is chosen as $Q = 3 H \xi (1+w_x) \rho_x$, thus, the interaction function depends on the energy density of the dark energy ($\rho_x$) and pressure like contribution ($p_x=w_x\rho_x$) of dark energy. This kind of interaction function is very efficient to successfully remove the singularity in the pressure perturbation equation (\ref{eq:deltap}) of dark energy at $w_x=-1$.   

For a clear understanding on how the dynamical $w_x$ in (\ref{eq:ZG-II}) effects the cosmological evolution in presence of interaction in the dark sector, we have also studied the non-interacting scenario with the same $w_x$ given in (\ref{eq:ZG-II}).    
In order to fit both the models with the observational data we use different combinations of recently available observational datasets. For the interacting scenario driven by the one parameter $w_x (z)$ model, given in equation (\ref{eq:ZG-II}), the coupling parameter ($\xi$) is obtained to be very small and consequently the interaction is less significant. It is also clear from Fig.~\ref{fig-one-parameter-interaction} that the coupling parameter $\xi$ is almost uncorrelated with other parameters. The present value of dark energy equation of state parameter strongly remains in the phantom regime. However, the analyses also show that within 68.3\% CL, the combined datasets CMB $+$ BAO $+$ HST and CMB $+$ BAO $+$ WL are in tension with the datasets CMB $+$ BAO $+$ RSD and CMB $+$ BAO $+$ RSD $+$ HST $+$ WL $+$ JLA $+$ CC (Fig.~\ref{fig-one-parameter-interaction}). We perform similar analyses with the non-interacting scenario (see Table \ref{tab:results-one-parameter-without-interaction}) and compared both the scenarios in  Fig.~\ref{fig-comparison-oneparameter} (for the full dataset CMB $+$ BAO $+$ RSD $+$ HST $+$ WL $+$ JLA $+$ CC) which shows that for this one parameter $w_x (z)$ model, the observational constraints, obtained in interacting and non-interacting scenarios, are almost indistinguishable. 

Now, from the effects of the interaction on the large scales of the universe, shown in Fig.~\ref{fig-cmb-power} (CMB spectra) and Fig.~\ref{fig-matter-power} (matter power spectra), we have a couple of observations.  From  
the power-spectra of the anisotropy of CMB temperature (Fig.~\ref{fig-cmb-power}), the power-spectra of matter density contrast (Fig.~\ref{fig-matter-power}) and also from the corresponding residuals  with respect to the base $\Lambda$CDM (right panels of both Fig.~\ref{fig-cmb-power} and Fig.~\ref{fig-matter-power}), it is clearly found that the interacting one parameter $w_x$ model sharply deviates from the base $\Lambda$CDM power-spectrum.

The direction of energy flow is another important quantity in this context that we have explored. For the present model, since the value of the coupling parameter ($\xi$) is essential for the stability of the interacting scenario (see the discussions after equation \ref{doom} of section \ref{sec-2}), thus, the direction of the energy flow between the dark fluids mainly depends on the factor $(1+w_x)$, that means, the direction of energy flow is affected by the nature of dark energy (for $w_x> -1$, $Q >0$ while for $w_x< -1$, we have $Q <0$).  A general behaviour of the interaction function has been shown for different values of $\xi$ (Fig. \ref{fig:Q-xi}) as well as for different values of $w_0$ with some fixed $\xi$ (Fig. \ref{fig:Q-w}). Now, from the constraints on $w_0$, one can see that the present analyses allow both quintessence ($w_0>-1$) and phantom ($w_0<-1$) regimes, and the mean values of $w_0$ remains in phantom one. Thus, the interaction rate at present,  $Q (z= 0)$ is allowed to have both positive and negative value, though $Q (z= 0)<0$ is slightly preferred. From the conservation equations (\ref{cons-dm}) and (\ref{cons-de}), it is clear that a negative value of the interaction function $Q$ indicates the energy flow from the dark energy to dark matter. 

As already mentioned that the present results clearly show that the dark energy equation of state parameter and the coupling parameter ($\xi$) of the interaction function are almost uncorrelated. It indicates that the dark energy equation of state parameter is in general degenerate with the possible interaction in the dark sector. In an earlier work \cite{Kunz:2007rk}, the author has discussed about the degeneracy in generalized dark energy models with respect to different cosmological probes and concluded that interacting dark energy is always equivalent to a class of non-interacting dark energy. From the present analysis, it can be particularly concluded that the dark energy equation of state is not distinguishable for interacting and non-interacting scenarios based on the present cosmological data at background and at linear perturbation level. Lastly, from the Bayesian analysis, we find that $\Lambda$CDM is still strongly favored over the IDE model. This might be the case related to the increased dimension of the IDE parameter space  compared to the 6-parameters based $\Lambda$CDM model.

\section*{Acknowledgments}
The authors are thankful to the referee for some important comments and suggestions to improve the work. W. Yang's work is supported by the National
Natural Science Foundation of China under Grants No.  11705079 and No.  11647153. EDV acknowledges support from the European Research Council in the form of a Consolidator Grant with number 681431. The authors thank  Rafael C. Nunes and N. Tamanini for several discussions.


\end{document}